\documentclass[fleqn,usenatbib]{mnras}

\usepackage{newtxtext,newtxmath}

\usepackage[T1]{fontenc}

\DeclareRobustCommand{\VAN}[3]{#2}
\let\VANthebibliography\thebibliography
\def\thebibliography{\DeclareRobustCommand{\VAN}[3]{##3}\VANthebibliography}

\usepackage[usenames,dvipsnames]{color}
\usepackage{graphicx}	
\usepackage{amsmath}	
\usepackage{enumerate}
\usepackage{enumitem}
\usepackage{siunitx}
\usepackage{CJK}
\usepackage{anyfontsize}

\newcommand{\Sref}[1]{Section \ref{#1}}

\sfcode`\.=1001\sfcode`\?=1001\sfcode`\!=1001
\newcommand{\Fref}[1]{\ifhmode \ifnum\spacefactor=1001 Figure \ref{#1}\else Fig.\ \ref{#1}\fi \else Figure \ref{#1}\fi}
\newcommand{\Eref}[1]{\ifhmode \ifnum\spacefactor=1001 Equation (\ref{#1})\else equation (\ref{#1})\fi \else Equation (\ref{#1})\fi}
\newcommand{\Teffnom}{\TextOrMath{$T^_{\mathrm{eff}\odot}$}{T_{\mathrm{eff}\odot}}}
\newcommand{\Teff}{\ensuremath{T_{\mathrm{eff}}}}

\newcommand{\EPIC}{\textsc{epic}}
\newcommand{\SAMD}{\textsc{samd}}
\DeclareSIUnit\angstrom{\mbox{\normalfont\AA}}

\title[Evolution of Sun-like stars]{Chemodynamic evolution of Sun-like stars in nearby moving groups}

\author[C. Lehmann et al.]{Christian Lehmann,$^{1,2}$\thanks{E-mail: christian.lehmann@geol.lu.se}
Michael T. Murphy,$^{1}$
Fan Liu (刘凡),$^{1,3,4}$
Chris Flynn,$^{1,5}$
\\
$^{1}$Centre for Astrophysics and Supercomputing, Swinburne University of Technology, Hawthorn, Victoria 3122, Australia 
\\
$^{2}$Lund Observatory, Department of Geology, S\"{o}lvegatan 12, SE-22362 Lund, Sweden
\\
$^{3}$School of Physics and Astronomy, Monash University, Melbourne, VIC 3800, Australia
\\
$^{4}$ARC Centre of Excellence for Astrophysics in Three Dimensions (ASTRO-3D), Canberra, ACT 2611, Australia
\\
$^{5}$OzGrav: ARC Centre of Excellence for Gravitational Wave Discovery, Hawthorn, VIC 3122, Australia
}

\date{Accepted XXX. Received YYY; in original form ZZZ}

\pubyear{2024}

\begin{document}
\begin{CJK*}{UTF8}{gbsn}
\label{firstpage}
\pagerange{\pageref{firstpage}--\pageref{lastpage}}
\maketitle

\begin{abstract}
Sun-like stars are well represented in the solar neighbourhood but are currently under-utilised, with many studies of chemical and kinematic evolution focusing on red giants (which can be observed further away) or turn-off stars (which have well measured ages). Recent surveys (e.g.\ GALAH) provide spectra for large numbers of nearby Sun-like stars, which provides an opportunity to apply our newly developed method for measuring metallicities, temperatures, and surface gravities -- the $\EPIC$ algorithm -- which yields improved ages via isochrone fitting. We test this on moving groups, by applying it to the large GALAH DR3 sample. This defines a sample of $72{,}288$ solar analogue targets for which the stellar parameter measurements are most precise and reliable. These stars are used to estimate, and test the accuracy and precision of, age measurements derived with the \SAMD\ isochrone fitting algorithm.
Using these ages, we recover the age--metallicity relationships for nearby ($\lesssim\SI{1}{kpc}$) moving groups, traced by solar analogues, and analyse them with respect to the stellar kinematics. In particular, we found that the age--metallicity relationships of all moving groups follows a particular trend of young ($\textrm{age}<\SI{6}{Gyr}$) stars having constant metallicity and older ($\textrm{age} \ge\SI{6}{Gyr}$) stars decreasing in metallicity with increasing age.
The Hercules stream carries the highest fraction of metal-rich young stars ($\sim\SI{0.1}{dex}$) in our sample, which is consistent with a migrating population of stars from the inner Galaxy, and we discuss the possible causes of this migration in the context of our results.
\end{abstract}

\begin{keywords}
instrumentation: spectrographs -- methods: data analysis -- stars: solar-type -- stars: fundamental parameters -- techniques: spectroscopic
\end{keywords}



\section{Introduction}
Stars similar to the Sun are important probes for the chemical and kinematic evolution of the local solar neighbourhood as they are numerous and their spectra are close to the spectrum of the Sun, which is well understood. 
In previous works, we developed an efficient algorithm to measure stellar parameters for stars similar to the Sun \citep[][]{CLehmann2022}, which was used to identify stars as solar analogues. The method we developed, $\EPIC$\footnote{\url{https://github.com/CLehmann94/EPIC}}, has reduced the uncertainties for $\Teff$, $\mathrm{[Fe/H]}$, and $\log g$ by a factor of 2--4 with respect to comparable previous methods when applied to stars very similar to our Sun.
We identified 72{,}288 solar analogue stars observed as part of the third data release of the GALAH survey \citep[hereafter GALAH DR3;][]{Buder2021} with the High Efficiency and Resolution Multi Element Spectrograph (HERMES) at the $\SI{3.9}{\metre} $ Anglo-Australian Telescope (AAT).  Where previously this has been used to search for solar twins at smaller Galactocentric radii than the Sun \citep{Lehmann2023}, here we apply it to the large spectral data set of GALAH DR3.
With our measurements of stellar parameters, and in combination with the measurement of kinematic properties from GALAH and Gaia, we probe the chemical and kinematic evolution of solar analogue stars focusing on their identification as part of moving groups to better understand the evolution of the Milky Way \citep{Freeman2002, Schoenrich2009, Loebman2011, Minchev2013, Minchev2015, Bensby2014}.

We track the evolution of stars via age measurements, which are themselves a unique challenge in stellar astrophysics. The only star with precise age measurements is the Sun ($t_\textrm{seis} = 4.57 \pm \SI{0.11}{Gyr}$, \citealt{Bonanno2002} and $t_\textrm{iso} = 4.5673\pm \SI{0.0003}{Gyr},$ \citealt{Connelly2012}) as its proximity allows for the applications of methods not possible for other stars, e.g.\ the analysis of solar seismic models with solar measurements or probing isotope element in the solar system.
Measuring ages of other stars is complicated, with direct measurements (similar to those for the Sun) being extremely difficult, so methods using chemical clocks \citep{Nissen2016, Aguirre2018, Casali2020, Hayden2022, Moya2022, Vazquez2022}, nucleocosmochronometry \citep{Cowan1999, Frebel2008, Shah2023} and asteroseismology \citep{Aguirre2018, Basu2018, Valentini2019} have been used to measure ages. These indirect measurements are themselves challenging and the age precision is often limited to $\sim\SI{1}{Gyr}$. In this work, we make use of the isochrone fitting method (used as early as in \citealt{Demarque1964} and later in many subsequent works, e.g. \citealt{Joergensen2005, Mints2017, Sanders2018, Rocha2022}). This uses stellar parameters of stars -- such as temperature, surface gravity, metallicity, and distance -- to find which age isochrone within the model match them best (within this work we use PARSEC isochrones, \citealt{Nguyen2022}).
Isochrone fitting is most effective (i.e.\ lowest relative uncertainties) when applied to sub-giant branch and turn-off stars, but can also provides results with relatively low uncertainties for stars like the Sun. The driving factors of uncertainty in this method are (i) the isochrone sampling and accuracy (i.e. the model) and (ii) the stellar parameters, and their uncertainties, that are used to determine the age.

Previously, isochrone fitting methods on Sun-like stars were not feasible with moderate-resolution spectra at modest $S/N$, as their relatively high stellar parameter uncertainties lead to age uncertainties \mbox{$\sim2$--$\SI{4}{Gyr}$}.
Nevertheless, isochrone fitting has been used on solar twins where precise stellar parameters were available from high resolution and $S/N$ spectra, especially on surveys with a small number ($<1{,}000$) of stars. \citet{Nissen2016} made use of isochrones to measure ages for solar type stars as their stellar parameter uncertainties were remarkably low [i.e.\ $\sigma\left(T_\textrm{eff}\right) = \pm \SI{6}{\kelvin}$ and $\sigma\left(\log g \right) = \pm \SI{0.012}{dex}$], which translates to $\sigma(\textrm{age}) = \SI{0.4}{Gyr} - \SI{0.8}{Gyr}$. Of course, this was made possible by high resolving power and high $S/N$ in their observed spectra, but such high age precision suggests many possibilities, such as better capacity to probe the star formation history of the Milky Way, among many others.

In this work, we derive ages of Sun-like stars from our own precise stellar parameters to explore the evolution of moving groups, i.e.\ with similar kinematic properties focusing on angular momentum. There are a number of moving groups that can be identified in the solar neighbourhood, most notable are the Hyades \citep[known since ][]{Proctor1869}, Horn, Sirius and Hercules streams \citep[][]{Eggen1995, Eggen1996, Fragkoudi2019}. These moving groups were better characterised in the projects of ESA's HIPPARCOS mission \citep[][]{Skuljan1999, Famaey2005, Famaey2007}, and later in the Gaia mission and the Gaia data releases \citep{Collaboration2016, Collaboration2018, Collaboration2021, GaiaCollaboration2022}.
Stars within these moving groups can be identified by their unique angular momentum as well as other factors like (in a Galactocentric coordinate system) azimuthal velocity, radial velocity, and radius. 
The recent work of \citet{Lucchini2022} attempted to identify these moving groups to better understand the structure of the Milky Way. They found that Hercules has a higher stellar density further towards the Galactic Centre while Sirius has a higher stellar density further away from it, which is most likely connected with their origin.

Most works in regards to moving groups focus on the Hercules stream as it transports stars from inner regions of the Milky Way towards the solar neighbourhood, which provides the unique opportunity to probe this different stellar population. Migration is thought to occur through multiple channels, i.e. `blurring' which forces stars on eccentric orbits where the outermost and innermost Galactocentric radii can be very different, and `churning' which changes the Galactocentric radius of a star without influencing its eccentricity (see \citealt{Sellwood2014} for a review). 
\citet{Dehnen2000} and \citet{Gardner2010} identified different moving groups as velocity patterns within the local Galactic disk, i.e.\ moving groups are identified by a combination of Galactocentric radial and azimuthal velocity that have a high density of stars in parameter space. \citet{Gardner2010} identified the Hercules stream at the Galactocentric radial velocity of $v_R=\SI{-50}{\kilo\metre\per\second}$ (i.e.\ moving toward the Galactic Centre) and the azimuthal velocity in the Galactic disk plane of $v_T=\SI{-30}{\kilo\metre\per\second}$ \citep[i.e.\ relatively slower than the Sun, ][]{Gardner2010a} as well as the Pleiades, Sirius and Hyades streams at different velocities. 

The Hercules stream is a moving group of stars in the solar neighbourhood and believed to have been caused by the Galactic bar. One possible cause would be the bar's outer Lindblad resonances (OLR). \cite{Monari2016} claim that the influence of a long and slow--rotating Galactic bar cannot explain the formation of the Hercules stream via 4:1 OLR, while a fast--rotating and short bar is able to reproduce the stellar density of the Hercules stream (and other moving groups) in velocity phase space.
However, \citet{Hunt2018} have shown that a more complex potential for the Galactic bar leads to the 4:1 OLR, reproducing a stream like Hercules with a long, slow-rotating Galactic bar.

There are alternative explanations for the Hercules stream's origin, such as \citet{DOnghia2020} where a large number of stars was initially trapped in the Lagrangian point located $90^\circ$ off the major axis of the bar and eventually escaped as the bar grows. This hypothesis is similar to \citet{PerezVillegas2017} who suggested that stars within the Hercules stream orbit the Lagrangian point caused by a large and slow Galactic Bar.
The result of both of these would be that a large number of stars follow similar orbits, which could then be observed as a stream. As this hypothesis would use the bar's corotation, it opposes the OLR origin. 
Here, we use solar analogue stars with the kinematic signature of the Hercules stream to find evidence for either of these origins using our accurate stellar parameters and ages calculated with $\EPIC$ and the `algorithm for estimating the sample age-metallicity distribution' (\SAMD, \citealt{Sahlholdt2020}). 

This paper is structured as follows. In \Sref{sec:selection_ch4} we define the selection of solar analogues for this study and in \Sref{sec:iso_age_ch4} we test the age measurements for our sample against those of GALAH DR3.
In \Sref{sec:Herc} we identify stars as part of moving groups in the solar neighbourhood and analyse them in terms of both age and metallicity. \Sref{sec:conclusion_ch4} contains the main conclusions and takeaways from this work.

\section{Stellar parameters and selection of stars}\label{sec:selection_ch4}

\subsection{\EPIC\ stellar parameter estimates}\label{ssec:EPIC}
All of the stars we use were taken from GALAH DR3 \citep{Buder2021}\footnote{GALAH DR3 is hosted at \url{https://datacentral.org.au/services/download/}}, with a total of $588{,}307$ stars in the catalogue. We use the following catalogues: Recommended catalogue of stellar parameters and abundances; Ages, masses, distances and other parameters estimated by BSTEP; and Galactic kinematic and dynamic information\footnote{Catalogues can be downloaded at \url{https://www.galah-survey.org/dr3/the_catalogues/}}.
We applied the $\EPIC$ algorithm \citep{CLehmann2022} to the spectra of all of these stars to measure their stellar parameters: temperature $\Teff$, surface gravity $\log g$ and metallicity [Fe/H]. The algorithm measures equivalent widths of a selected set of absorption lines within spectra. It is first applied to a grid of spectra, with a range of stellar parameters, to construct a model connecting the equivalent width of individual lines to the stellar parameters of the stars. When the algorithm is then applied to a target star in our sample (from GALAH DR3 in this case), it measures the absorption line equivalent widths and then inverts the stellar parameter model to measure temperature, surface gravity and metallicity. Crucially, all equivalent width measurements are conducted in a differential way relative to a reference spectrum -- in this case, that of the Sun. This line-by-line differential approach reduces systematic errors in the equivalent width variations between stars and, ultimately, the stellar parameters. Note that the stellar parameters within GALAH DR3 were measured with Spectroscopy Made Easy \citep[SME][]{Piskunov2017}, as described in \citet{Buder2021}.

\begin{figure}
	\centering
	\includegraphics[width=0.4\textwidth]{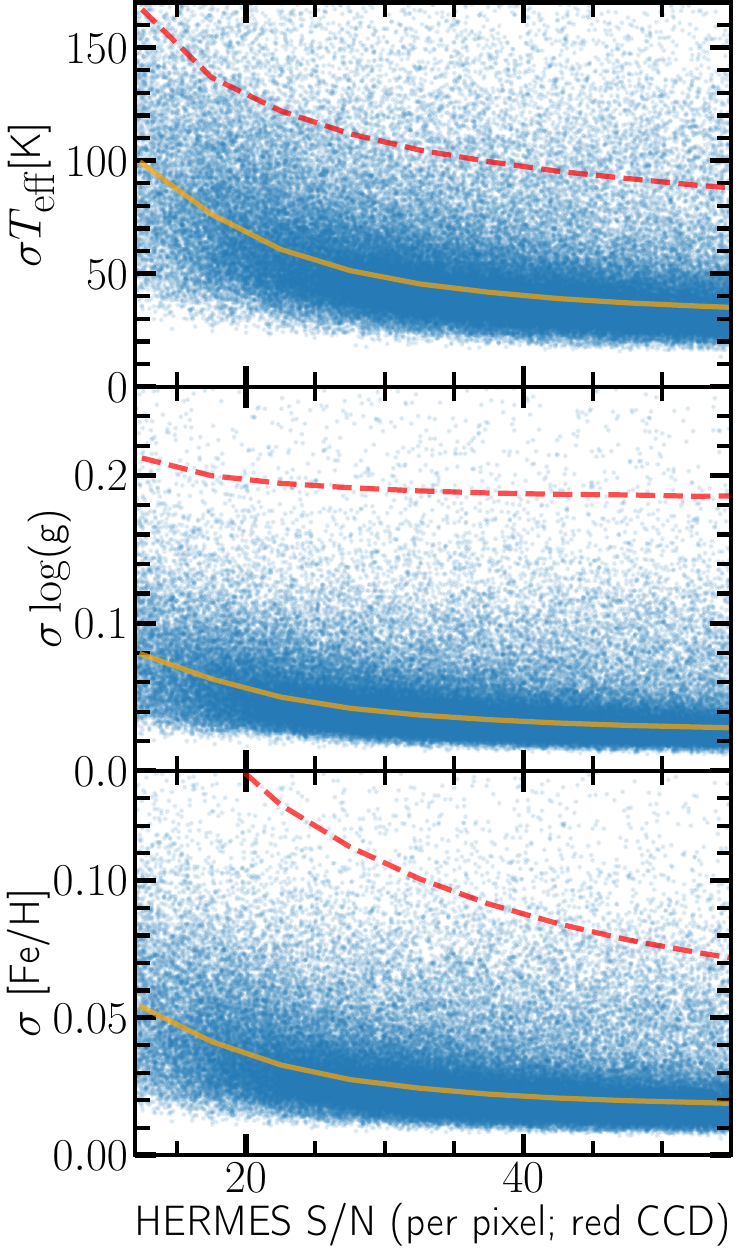}
	\caption[The uncertainties for the three main stellar parameters against the $S/N$ in the red CCD]{The uncertainties for the three main stellar parameters $\Teff$, $\log g$ and [Fe/H] against the $S/N$ in the red CCD for our selection of GALAH target stars. The blue dots represent \EPIC\ measurements with the orange line as a running uncertainty median. The red dashed line represents the same median for GALAH DR3 measurements.}
	\label{fig:uncertainties_DR3}
\end{figure}

Each of these parameters have improved uncertainties compared to the measurements of the GALAH survey, down by a factor of 2--4. 
In $\EPIC$, the stellar parameter estimates are derived by fitting a model of how the equivalent widths of a sample of spectral lines vary with stellar parameters against the measured equivalent widths from a spectrum. The uncertainties in the stellar parameters derive from deviations between the model and the measured equivalent widths as well as the statistical (photon) noise in the spectral lines (see section 3.5 of \citet{CLehmann2022}).
We visualise our measured uncertainties in \Fref{fig:uncertainties_DR3}, where we can see the improvement of \EPIC\ stellar parameters compared to those of GALAH DR3. The running median uncertainty for all three measured parameters is down by a factor of >2 at all $S/N$ values. This is consistent with similar findings in \citet[][figure 10]{CLehmann2022} from a much smaller sample ($\sim$2000 spectra) drawn from GALAH DR2. 
We note here that, strictly speaking, the accuracy of the stellar parameters from $\EPIC$ is difficult to evaluate. Because $\EPIC$ uses a stellar library of composite GALAH spectra to establish models of spectral line strengths with variations in stellar parameters, its resulting stellar parameters necessarily share the same calibration as the GALAH stellar parameters and, therefore, their overall accuracy. This is discussed extensively in \citet{Lehmann2023} where, for the purposes of that paper, we re-calibrated the EPIC stellar parameters to those established from detailed measurements of a smaller sample of higher-resolution solar twin spectra. However, in this paper we maintain the original GALAH calibration of $\EPIC$'s stellar parameters because isochrone age measurements make use of stellar parameters calibrated similarly.

With this calibration, we compared our stellar parameters with those from APOGEE DR17, where $7{,}890$ stars were overlapping with our sample. This was done to better determine the accuracy of $\EPIC$ against another large spectroscopic data set.
For these $7{,}890$ stars specifically, [Fe/H] is in good agreement, while both $\Teff$ and $\log g$ seem to have a constant offset of $\sim\SI{100}{\kelvin}$ and $\sim\SI{0.1}{dex}$ respectively. These constant offsets are not concerning, as the results within each analysis have been interpreted independently from each other to arrive at similar conclusions. This means that $\EPIC$ has relatively good accuracy, even when tested against other data sets and analyses.

\subsection{Solar analogue selection}\label{ssec:analogue_selection}
We select solar analogue targets using only their temperature and surface gravity, and retain the metallicity as a free parameter we can use in this study. Therefore, the following stellar parameter conditions were chosen:
\begin{align}
    \text{Solar analogue selection} &\rightarrow \left\{
    \begin{array}{r@{\,}l}\label{eq:SA_selection}
        \Teffnom&\pm\,\SI{300}{\kelvin},\\
        \log{g_\odot}&\pm\,\SI{0.4}{dex},\\
    \end{array} \right.
\end{align}
with $\Teffnom=\SI{5772}{\kelvin}$ and $\log{g_\odot}=\SI{4.44}{dex}$ \citep{Prsa2016}. This leaves us with $166{,}572$ potential solar analogues.

The line broadening of these targets is determined via the method explained in \citet[][section 3.2]{Lehmann2023}.
The broadening measurements are important to identify stars for which the effectiveness of the $\EPIC$ algorithm is not guaranteed. This is because the $\EPIC$ algorithm assumes no additional broadening terms for lines, allowing it to use a fixed wavelength interval around lines to measure the equivalent width \citep[][section 2.4]{CLehmann2022}. A broadened line could be wider than the width of this measurement window and therefore be inaccurately measured. Broadened lines can also cause additional blending between some absorption lines. These factors mean that the \EPIC\ stellar parameter model is not ideal to predict parameters from these measurements.
Therefore, we use the same condition that was used in \citet{Lehmann2023} for the selection of viable probes to measure the fine-structure constant $\alpha$: $\textrm{line broadening} <\SI{6}{\kilo\metre\per\second}$ (leaving $162{,}858$ stars).

While the stellar parameter measurements of $\EPIC$ have been verified to identify solar analogue stars with good precision, we apply some additional selection criteria to eliminate potential outliers, i.e.\ stars that are less solar-like than expected.
First, we apply a colour cut using photometry from the 2MASS catalogue (provided within the GALAH DR3 catalogue). We limit the 2MASS colours to $0.2 < (J - K) < 0.7$ which roughly corresponds to the solar analogue temperature range. This provides a photometric criterion which is independent of our spectroscopic selection criteria to assist in removing outliers. For this to be applied stably, we ensure a low reddening, i.e.\ E$(B-V)<0.2$\,mag, from the \citet{Schlegel1998} dust map. This leaves us with $152{,}652$ stars. 
For the distance, which are calculated via parallax measurements with Gaia \citep{GaiaCollaboration2022}, we require a relative uncertainty less than $20\%$ to ensure that our estimates of metallicity and age variation with distance in \Sref{sec:Herc} are reliable. This cut leaves $149{,}711$ stars. We also require low uncertainties for stellar parameters measured with $\EPIC$, i.e.\ $\sigma_{T\textrm{eff}} < \SI{100}{\kelvin}$ and $\sigma_\textrm{[Fe/H]} < \SI{0.1}{dex}$. This ensures that $\EPIC$ found a reasonable match between the stellar parameter model and equivalent width measurements in these stars.
Overall, these quality control selections eliminate another $\approx18\%$ of the stars left from line-broadening cut, resulting in $132{,}990$ viable target stars.

Furthermore, we want to select stars that are primarily associated with thin disk dynamics in the Milky Way. This is done because we are mainly interested in the dynamics and moving groups that resemble stars within the thin disk, while the thick disk (or stellar halo) would need additional care with regards to their moving groups. 
We chose to use $z_\mathrm{max}<\SI{500}{pc}$ (the amplitude of the orbit in $z$-direction) for this, which leaves $72{,}288$ stars.  $z_\mathrm{max}$ is a good indicator of which disk population stars are a part off as it takes into account the whole orbit of a star, as opposed to e.g.\ $z$, which only considers the current $z$-position.

\section{Isochrone age measurements for Sun-like stars}\label{sec:iso_age_ch4}
In this section we derive the ages of the target stars with a Bayesian isochrone fitting approach called \SAMD\ \citep[developed and tested in][]{Sahlholdt2020, Sahlholdt2021, Sahlholdt2022}, where we use $\EPIC$ stellar parameters (temperature, surface gravity, metallicity) as well as the parallax and $G$ band photometry (from Gaia) as inputs.
Using PARSEC isochrones \citep{Nguyen2022}, \SAMD\ calculates a probability density function (PDF) for the ages of stars. It is debatable if the median or mode of the PDF should be used as the most probable age, which is also dependent on the exact use, e.g.\ \Fref{fig:Age_diff_GALAH} uses the mode to compare these measurements with GALAH's BSTEP because they used a similar metric.
The PDF also lets us define the uncertainty as half the difference of the 16th and 84th percentile, which should be a good approximation for the real uncertainty. The validity of the absolute values for ages in this work are debatable, dependent on the isochrones chosen and also the calibration of stellar parameters. However, it is most important that the population is comparable with itself so that age trends can be accurately presented. 
For the sake of comparison in \Sref{sec:Herc}, we make use of another algorithm \citep[q$^2$, ][]{Ramirez2014} with another set of isochrones \citep[Yonsei-Yale, ][]{Demarque2004}. We do this to show that we arrive at the same qualitative results independently from the exact choice of isochrones while leaving the discussion about isochrones accuracy and applicability to other studies \citep[e.g.][]{Rottensteiner2024}.

We test our approach with two metrics: (i) How well our age estimates compare with those from previous studies for the same stars?: We test our ages against those those derived with the GALAH DR3 BSTEP method \citep{Sharma2017} using the same spectra;
(ii) How do our stellar age uncertainties compare to those derived from the GALAH DR3 BSTEP method using the same spectra?
Another possible test could be to compare our age estimates with studies using higher-resolution spectra of stars common with the GALAH sample. Unfortunately, GALAH focuses on stars fainter than those typically used in such high-resolution studies, so this is not an option.
However, in \citet{CLehmann2022} we used EPIC to derive stellar parameters for the stars studied with high-resolution HARPS spectra by \citet{Nissen2016} and \citet{Casali2020}. While these stars were not observed by GALAH, we simulated GALAH spectra by degrading the resolution and signal-to-noise ratio of the HARPS spectra. The EPIC stellar parameters compared very closely with those derived by the original authors from the HARPS spectra. We have used the methods in this paper to estimate ages from the EPIC-derived stellar parameters of these stars and, not surprisingly, they also compare very well.
When measuring the ages of stars using stellar parameters, we also considered if abundances that are not Fe (specifically $\alpha$ abundance) should influence the age measurement. 
In \citet{Sahlholdt2022}, an enhancement of $\alpha$ abundance was taken into account by a correction term found in equation (3) of their work. However, one possible weakness of the $\EPIC$ algorithm, i.e.\ different elemental absorption lines influencing the metallicity instead of just Fe, actually works in favour of the method here as it naturally takes into account enhancements in $\alpha$ that would need to be corrected for otherwise. As a means to test this, we applied the \citet{Sahlholdt2022} correction to our metallicities and compared the resulting ages with GALAH BSTEP ages, where we found worse agreement than without the correction applied. 


\subsection{GALAH BSTEP comparison}\label{sec:bstep}
The BSTEP approach implemented by the GALAH survey team \citep{Sharma2017} was designed to estimate ages for stars of all types within their sample. BSTEP is a Bayesian approach that attempts to find the most probable set of intrinsic parameters (including age) using all the available observable quantities (e.g.\ stellar parameters, photometric colours and location within the Galaxy).
It makes a number of key assumptions, namely that a star's mass, metallicity and distance is independent of its age. GALAH DR3's BSTEP method used a grid of stellar models which they constructed using PARSEC-v1.2S isochrones \citep{Chen2015}. In essence, this constructs another model of isochrones in a different parameter space, which can convert one set of stellar parameters into others like age or initial mass. In contrast, our approach focuses only on solar type stars. We employed the more precise stellar parameters derived from the $\EPIC$ method, differentially with respect to the Sun, as input to the age estimate using \SAMD\ with more up-to-date PARSEC-v2.0 isochrones.

\begin{figure}
	\centering
	\includegraphics[width=0.48\textwidth]{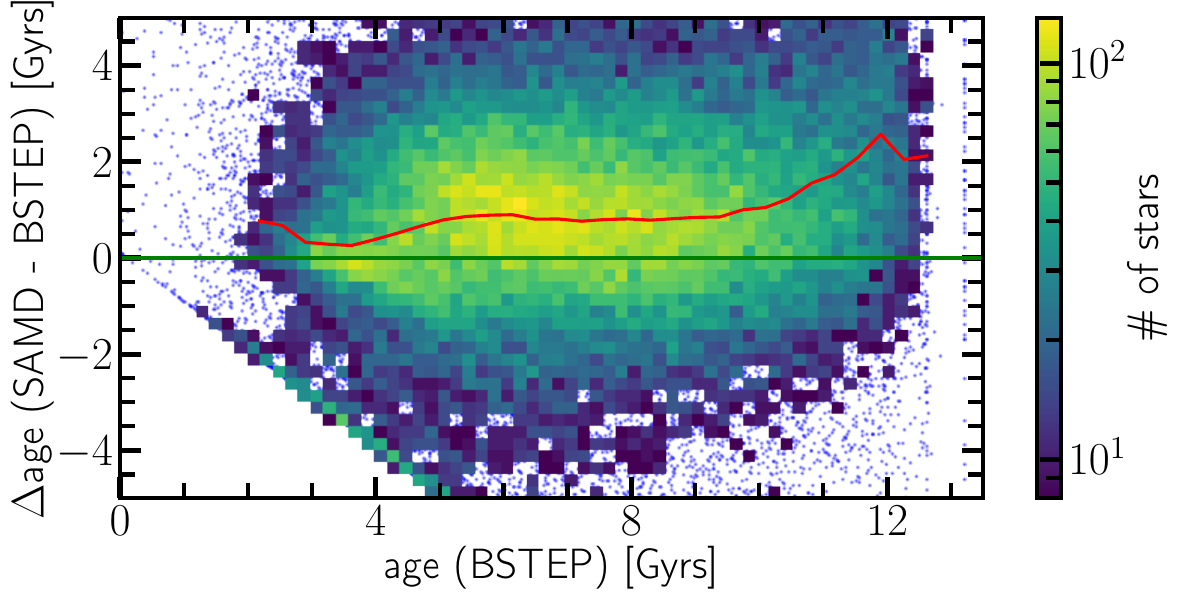}
	\caption[Age difference between our own $\EPIC$/\SAMD\ method and the GALAH BSTEP algorithm]{
    Age difference between the GALAH BSTEP and $\EPIC$/$\SAMD$ methods, as a function of the BSTEP ages, for our $72{,}288$ solar analogue stars. The number of stars in each pixel is represented in the colour map. The red line represents the median age difference and highlights the clear systematic difference observed between the two methods. For pixels with less than 8 stars, the data are represented as individual points. Note that in the \SAMD\ algorithm, ages larger than the age of the universe are permitted (though they have to be statistically consistent with the age of the universe). The unpopulated area in the bottom left of the figure is driven by \SAMD\ not allowing ages $<\SI{0}{Gyr}$.}
	\label{fig:Age_diff_GALAH}
\end{figure}

In \Fref{fig:Age_diff_GALAH} we show the differences between the BSTEP and $\EPIC$/\SAMD\ age estimates for our set of $72{,}288$ solar analogue stars as a function of BSTEP age.
There is a clear, systematic age difference between the two methods, highlighted by the running median in the figure (red solid line), of up to $\sim$\SI{2}{Gyr} for relatively old stars, with the $\EPIC$/\SAMD\ ages being generally older. 
The large discrepancy for older stars is most likely because \SAMD\ allows ages of up to $\SI{20}{Gyr}$, which spreads the older stars over a large age range, whereas BSTEP does not measure ages older than $\sim\SI{13.2}{Gyr}$, i.e.\ the age of the universe. To confirm this, we applied \SAMD\ to the GALAH stellar parameters and found that these systematic age differences all but disappear, i.e.\ the average difference reduces to $\lesssim\SI{0.2}{Gyr}$ between $4$ and $\SI{12}{Gyr}$.
Younger stars seem to have median differences of $\Delta\mathrm{age}=\SI{0.5}{Gyr}-\SI{1}{Gyr}$, which arises from a combination of updated isochrones and slightly different parameter calibrations. 
The larger age range also influences the measurement for the younger population, as this is the measurement is based on probabilities, i.e.\ younger stars will be measured as slightly older, possibly causing this trend we observe.
Nevertheless, the relatively small differences between age estimates, with only small trends with age, indicate that comparison between stars using the age estimates will likely allow for accurate analysis of relative age tends, which we are aiming for in this work.

\subsection{Age uncertainties}\label{sec:age_uncertainties_ch4}

\begin{figure}
	\centering
	\includegraphics[width=0.48\textwidth]{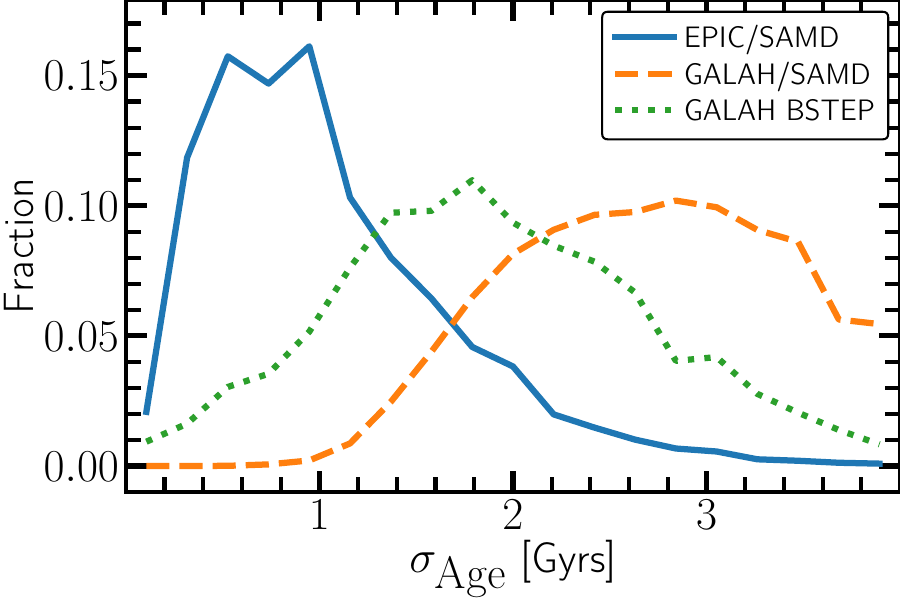}
	\caption[The distribution of uncertainties for age measurements of three different methods]{The distribution of uncertainties for age measurements of three different methods. The targets contributing to this diagram (for all three methods) are the same 72{,}288 solar analogue stars used throughout this paper. The blue line uses the Bayesian isochrone fitting method with PARSEC-v2.0 isochrones and applied to stellar parameters measured with the $\EPIC$ algorithm. The dashed orange line uses the same method but with stellar parameters measured by the GALAH survey. The dotted green line represents the BSTEP ages directly taken from the GALAH DR3 catalogue.}
	\label{fig:Age_uncertainties}
\end{figure}

The age uncertainties from isochrone fitting generally have main two sources: the stellar parameter uncertainties and the isochrone model uncertainties. The high precision of the input stellar parameters from \EPIC\ improves the age uncertainties by a factor of $\sim 2$ compared to GALAH BSTEP and a factor of $\gtrsim2.5$ compared to isochrone fitting with higher stellar parameter uncertainty (see below). This also means, that the isochrone physics and its inherit uncertainties become a more important factor for the age uncertainty. 
The contribution of these sources of uncertainty are difficult to quantify because both of them are necessary for the measurement of age.  However, one can expect there to be a lower limit for age uncertainties generated from the isochrone model itself. We measure the age using the median of the PDF, and the $1\sigma$ uncertainty as half the difference between the 16th and 84th percentile of the PDF. 

In \Fref{fig:Age_uncertainties} we show the resulting age uncertainties for our analysis with $\EPIC$/\SAMD\ for the selected $72{,}288$ solar analogue stars. To illustrate the improvement stemming directly from the more precise \EPIC\ stellar parameters, we applied the same \SAMD\ isochrone algorithm but with the GALAH DR3 stellar parameters as input (labelled `GALAH/\SAMD' in the figure). The age uncertainties decrease by a factor of $\approx$2.5 when using the \EPIC\ stellar parameters, from a range of $\sigma_\textrm{age}\sim 1.8-\SI{4}{Gyr}$ for GALAH/\SAMD\ to $\sim0.2-\SI{1.5}{Gyr}$ for most of the $\EPIC$/\SAMD\ results.
Additionally, \Fref{fig:Age_uncertainties} compares the $\EPIC$/\SAMD\ uncertainties with those from GALAH's BSTEP approach for the same stars, which are currently the only published age uncertainties for this large set of spectral data. These have uncertainties typically $\approx$1.5 smaller than the GALAH/\SAMD\ estimates we introduced above. While the EPIC/\SAMD\ uncertainties are substantially ($\approx$1.5 times) smaller than the GALAH BSTEP uncertainties, comparing them is not straightforward: BSTEP achieves average age uncertainties of $\sigma_\textrm{age}\approx$0.8-\SI{2.4}{Gyr}, substantially less than is possible with \SAMD\ using GALAH's stellar parameters (i.e.\ GALAH/\SAMD), because BSTEP incorporates a larger array of measurements to determine the age estimate (\Sref{sec:bstep}).

In future applications, it should be possible to combine the $\EPIC$ stellar parameters with the additional information considered in GALAH's BSTEP approach to obtain even more precise age estimates for solar-type stars in GALAH. Nevertheless, the considerable improvement in stellar ages demonstrated using \EPIC, even for main sequence, solar-type stars, should bring substantial benefit for studies of Galactic chemical and kinematic evolution when using solar-type stars. The next section considers one such case where the improved age uncertainties enable more reliable age--metallicity relationships to be studied for local moving groups using solar-like stars.

\section{Moving groups in the solar neighbourhood}\label{sec:Herc}
For our set of 72{,}288 identified solar analogue stars (\Sref{sec:selection_ch4}), we measured the stellar parameters effective temperature, surface gravity, iron metallicity and age. Additionally, GALAH provides kinematic properties of these stars within their catalogue \citep{Buder2021}, which are calculated with astrometric information from Gaia EDR3/DR3 \citep{Collaboration2021, Collaboration2023} and spectral radial velocity measurements (measured as a redshift/blueshift in the spectra compared with template spectra from \citealt{Zwitter2018}). For the distances, both the geometric and photogeometric distances from \citet{BailerJones2021} were used. For the calculations of orbital information, \textsc{GALPY} \citep{galpy15} was used.
This gives us the opportunity to determine if the selected solar analogues are part of any local groups of moving stars. In the literature, the moving groups that are most well-studied are Hyades, Horn, Hercules and Sirius (see \citealt{Antoja2010} for a review on the field).

We focus here on identifying stars as part of these moving groups via their angular momentum (\Sref{sec:identification_streams} below). Together with the $z_\mathrm{max}$ selection in \Sref{sec:selection_ch4}, this is quite restrictive on the particular orbits of these stars. When viewing this through the lens of stellar migration, similar angular momentum between stars implies a similar level of `blurring' (increase of eccentricity of a stellar orbit, e.g.\ via Lindblad resonance with the disk). So the trends we are seeing in this work are indicative of how effective blurring is for moving stars within the Milky Way.
The GALAH survey uses a second way (in addition to our selection via angular momentum) to identify stars as part of moving groups: measuring constant orbital energy \citep[][]{Khanna2019}. The origin of moving groups with constant orbital energy is less well understood, but it seems that most local moving groups [with the exception of Hercules and the Gaia Moving Group 1 (hereafter GMG1)] fall into this category.

\subsection{Identifying moving groups}\label{sec:identification_streams}
As stated above, the angular momentum of stars has been used to identify them as part of a particular stream. The angular momentum is defined as $L_z= v_\phi \times R$, where $v_\phi$ is the azimuthal velocity around the Galactic Centre within the Galactic disk and $R$ is the distance from the Galactic Centre. Alternatively, one could define this selection via guiding radius instead, which would be equivalent to this selection.
This selection especially aims at identifying streams that originate from large scale blurring events, which keep the angular momentum constant while introducing an increase in eccentricity into the orbits of stars. We made this choice to better investigate Hercules which likely originates within the Galactic bar \citep{Hunt2018}, which applies blurring to stars on larger orbits.

\begin{figure}
	\centering
    \includegraphics[width=0.48\textwidth]{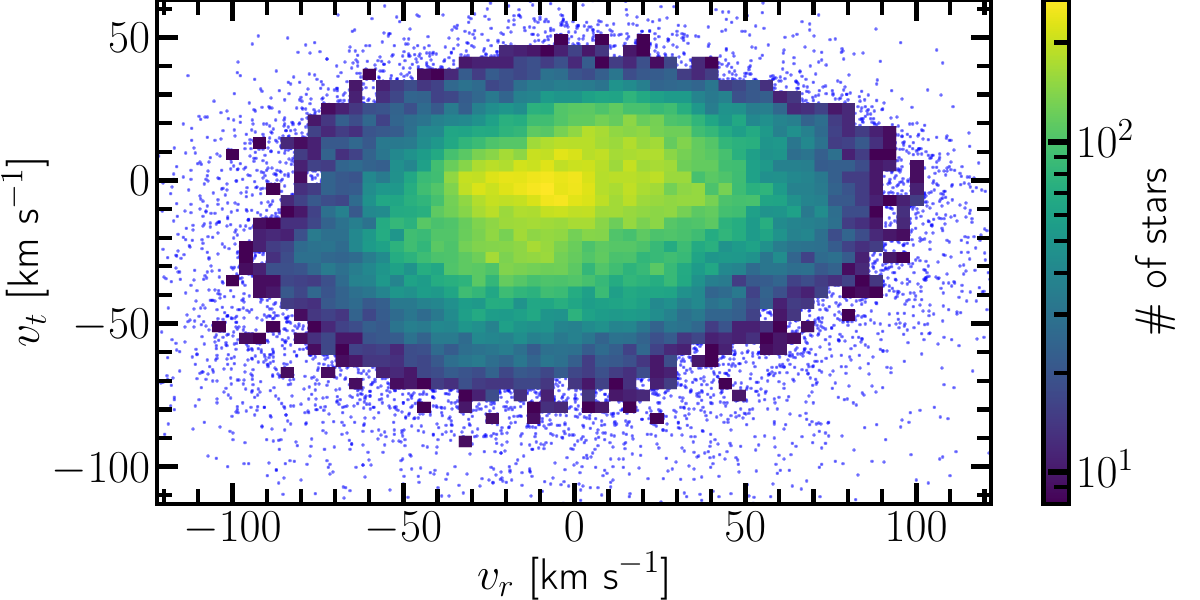}
	\caption[The velocity structure of our selected sample of nearby solar analogues]{The velocity structure of our selected sample of nearby solar analogues. The kinematic information is taken from the GALAH DR3 catalogue \citep{Buder2021}. The x-axis is the velocity in Galactocentric radial direction (where a negative sign means towards the Galactic Centre) while the y-axis is the velocity in the direction of rotation relative to the circular velocity of the solar neighbourhood \citep{Abuter2019}, where a negative sign means a star is slower than the average solar neighbourhood star. Rectangular bins with less than 8 stars in them, show individual stars. The large concentrations of stars within this diagram can be explained with moving groups as the feature at $(v_R, v_T) = (-5, 0)\,\SI{}{\kilo\metre\per\second}$ is a combination of Hyades and smaller moving groups, while the feature at $(v_R, v_T) = (-20, -30)\,\SI{}{\kilo\metre\per\second}$ is the largest component of what we identify as the Hercules stream. These velocities have been used to identify stars as parts of moving groups \citep[e.g.][]{Gardner2010}.}
	\label{fig:vR_vT}
\end{figure}

The moving groups we inspect are Sirius ($L_z = \SI{2050}{kpc\, \kilo\metre\,\second^{-1}}$), Hyades ($L_z = \SI{1880}{kpc\, \kilo\metre\,\second^{-1}}$), Horn ($L_z = \SI{1810}{kpc\, \kilo\metre\,\second^{-1}}$), Hercules ($L_z = \SI{1615}{kpc\, \kilo\metre\,\second^{-1}}$), GMG1 ($L_z = \SI{1385}{kpc\, \kilo\metre\,\second^{-1}}$) and Arcturus ($L_z = \SI{1210}{kpc\, \kilo\metre\,\second^{-1}}$), where $L_z$ was calculated using $v_\phi$ of the streams from \citet{Lucchini2022} and $R=R_\odot=\SI{8}{kpc}$. 
We have used a tolerance of $\SI{35}{kpc\, \kilo\metre\,\second^{-1}}$ around these values to identify stars as part of the stream, which was chosen to include a reasonable number of stars in each stream but also to keep each star distinct to belong to only one stream. This way of identifying stars is not very common in the literature, as normally moving groups are identified through concentrations of stars in kinematic phase space (e.g.\ \Fref{fig:vR_vT}). However, as we focus especially on the radial migration aspect via this selection, it seems to be reasonable. 
Both the angular momentum \citep{Lucchini2022} and kinetic energy \citep{Fragkoudi2019} were used in the literature with varying results. However, we ignore selection differences between these two methods as they are small within the solar neighbourhood due to relatively small variation of galactocentric radius $R$, but corrections might be introduced in future work.

\begin{figure}
	\centering
	\includegraphics[width=0.48\textwidth]{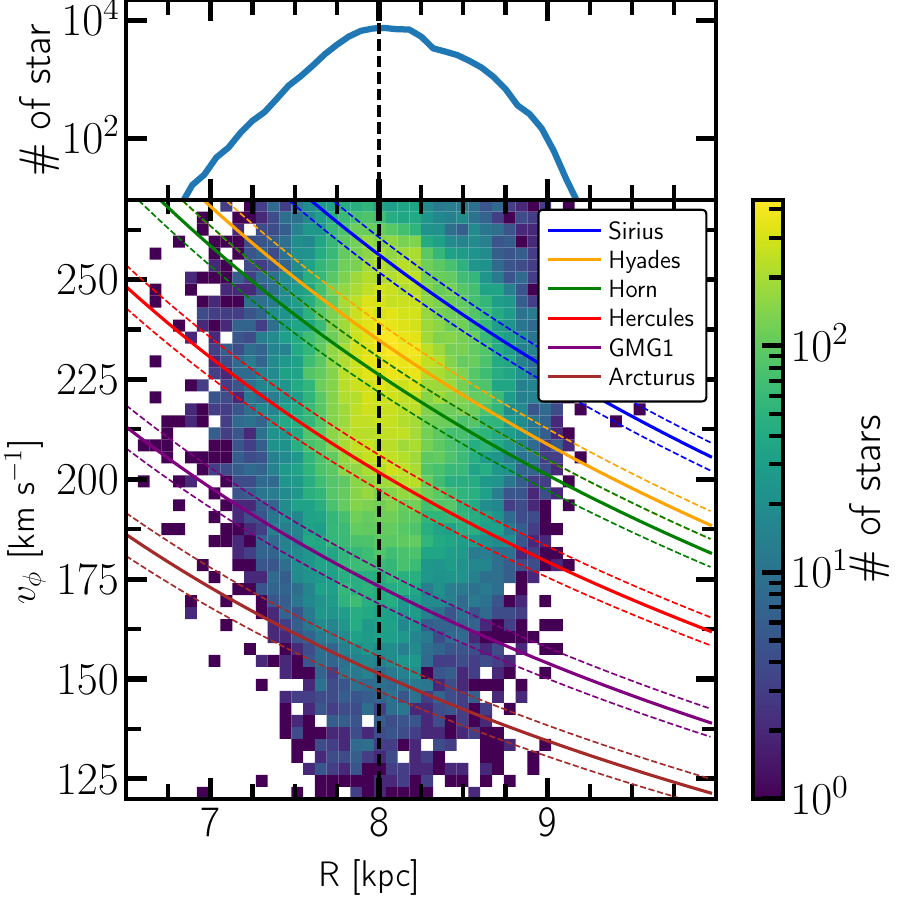}
	\caption[2D density map of the number of stars per pixel from our solar analogue sample on a azimuthal velocity-Galactocentric radius plane]{2D density map of our solar analogue sample on a azimuthal velocity-Galactocentric radius plane. Each of the solid coloured lines corresponds to the angular momentum of a specific stream as given in the legend. Note that these coloured lines follow constant angular momentum, which is not the ideal method of identifying some of these moving groups, but it is an adequate approximation when staying within the solar neighbourhood, here defined as $\lesssim\SI{1}{kpc}$ from the Sun. The dashed lines indicate the area in phase space that we use to identify stars as part of moving groups. The vertical dashed line indicates $R_\odot$ \citep{BailerJones2018}.}
	\label{fig:stream_identification}
\end{figure}

In \Fref{fig:stream_identification} we show a star density map of azimuthal velocity versus Galactocentric radius for our sample of solar analogue stars. We added lines with constant angular momentum and a tolerance around them which represent our stream selection. The bulk of stars in our sample are within $\SI{500}{pc}$ of the Sun with $\gtrsim100$ stars in each $\SI{70}{pc}$ bin for the range $\SI{7}{kpc} < R < \SI{9}{kpc}$.

\subsection{Age, metallicity and Galactocentric radius relationships within moving groups}\label{sec:age_metal_ch4}
\begin{figure*}
	\centering
	\includegraphics[width=0.8\textwidth]{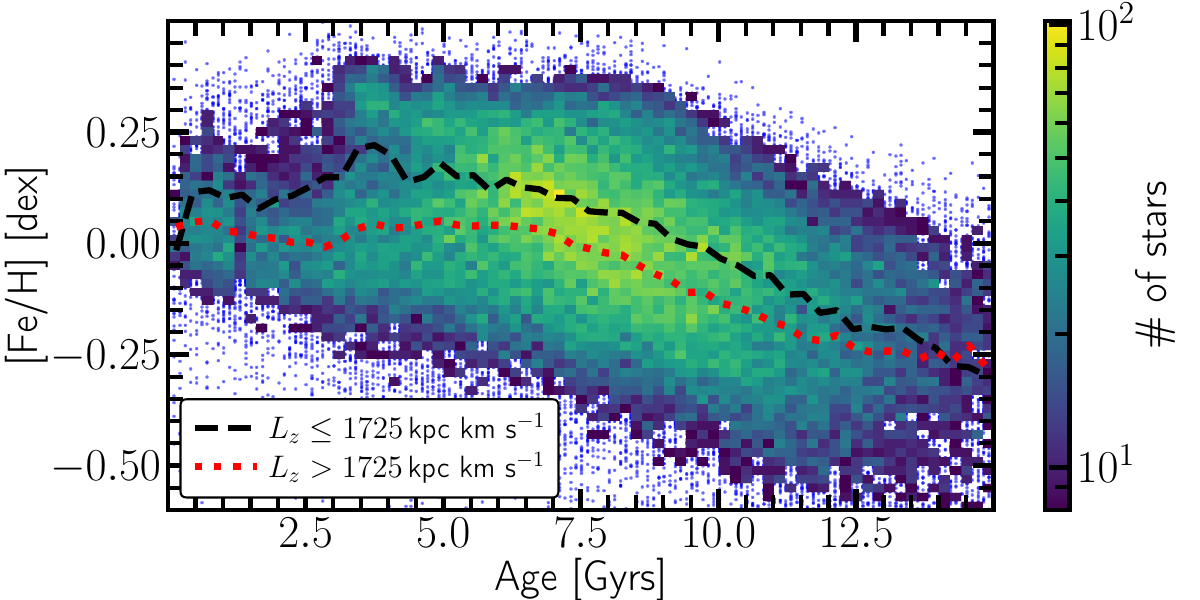}
	\caption[The age--metallicity distribution as a 2D colour map]{The age-metallicity distribution as a 2D colour map. The stars in this figure were analysed and selected according to the methods described in \Sref{sec:selection_ch4}. For rectangular pixels with less than 8 stars, individual stars are shown. The largest stellar population in the solar neighbourhood seems to be medium age ($\textrm{age} = 6-\SI{8}{Gyr}$) and somewhat metal-rich ($\textrm{[Fe/H]}\approx\SI{0.1}{dex}$).}
	\label{fig:age_metal_DR3}
\end{figure*}

Besides the identification of solar analogue stars, the main contributions of our work with GALAH spectra is the measurement of both the main spectroscopic stellar parameters and age of these stars with significantly lower uncertainties, of order $\sigma\left(\Teff, \log g, \textrm{[Fe/H]}\right) \approx \left(\SI{35}{\kelvin, \SI{0.03}{dex}, \SI{0.02}{dex}}\right)$ for spectra with $S/N>50$, and typical age errors $\sigma_\textrm{age}\approx\SI{1.0}{Gyr}$. 
As a comparison for analysis of individual moving groups below, we show the overall age--metallicity relationship for our whole sample of solar analogues in \Fref{fig:age_metal_DR3}. The figure also shows a high and a low angular momentum running median, showing the known trend \citep[e.g.][]{Willett2023} that stars with smaller angular momentum (black dashed line) have higher metallicities on average even at the same age.
The behaviour of stars overall is well known: older stars are metal-poor as the gas at the time of their birth reflects the star formation history of the Galaxy \citep[see e.g.][]{Minchev2018}.
Over the evolution of the Milky Way, supernova events of massive stars produce metals, which are funnelled back to the interstellar medium. This leads to an increase in metallicity for the interstellar medium and ultimately to more metal enriched stars born from its material. This holds true until medium aged stars ($\sim$3--\SI{6}{Gyr}), where we see a peak in the concentration of metal-rich stars ([Fe/H]$>\SI{0.2}{dex}$) with younger stars ($\lesssim\SI{3}{Gyr}$) having median metallicities around [Fe/H]$\approx\SI{0.0}{dex}$. 
This has multiple possible explanations, with an in-fall of fresh gas and a large timescale for stellar migration being common explanations in the literature \citep[e.g.][]{Sahlholdt2022}.

Another visible feature is that the concentration of stars in the medium age range is higher than for particularly young ($<\SI{4}{Gyr}$) and old ($>\SI{8}{Gyr}$) stars. The lower number of old stars can be explained by older solar analogue stars transitioning into red giants, which means that fever of them can be found in our sample than there were initially. Furthermore, there might be selection effects introduced both by the GALAH target selection and our own selection done in \Sref{sec:selection_ch4}.
The relatively low number of young ($<\SI{4}{Gyr}$) solar type stars we see might be because the main region of stellar formation is further towards the Galactic Centre and these stars need time to migrate outwards to the solar neighbourhood, where we can observe them. 
This could also partially explain the relatively low metallicity of the youngest stars (the population with $<\SI{2}{Gyr}$ has a median metallicity of $\approx\SI{0.0}{dex}$) as the observed stars might be dominated by locally born solar analogues at a Galactocentric radius of $\SI{8}{kpc}$. As a comparison, the majority of stars aged around $\SI{4}{Gyr}$ have measured metallicities between $\SI{0.0}{dex}$ and $\SI{0.3}{dex}$ in our sample.  
Since the local gas is less metal enriched than the inner Galaxy, we would expect locally born stars to have an overall lower metallicity distribution and only to reach solar metallicity relatively recently.

\subsubsection{Galactocentric radius-dependent age and metallicity changes}
\begin{figure*}
	\centering
	\includegraphics[width=0.85\textwidth]{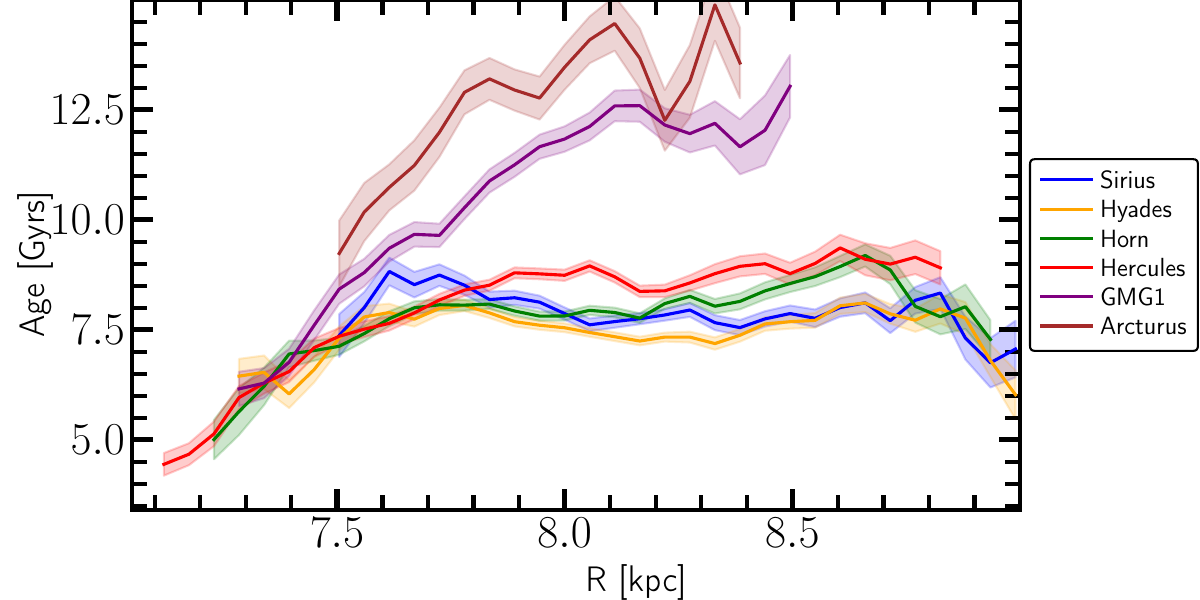}
	\caption[The median age of stars within the selected moving groups with respect to the Galactocentric radius]{The median age of stars within the selected moving groups with respect to the Galactocentric radius. 
    The bin width in Galactocentric radius is $\SI{0.1}{kpc}$ and the bins are overlapping. The 1-$\sigma$ statistical uncertainty (shown in the shaded areas) on the median lines is small, due to the large numbers of stars in each bin. We do not show bins containing fewer than 20 stars, which is why the mean relation does not always cover across all radii. On average, stars scatter by $\sim\SI{2.3}{Gyr}$ around their respective median lines.
    }
	\label{fig:stream_age}
\end{figure*}

\begin{figure*}
	\centering
	\includegraphics[width=0.85\textwidth]{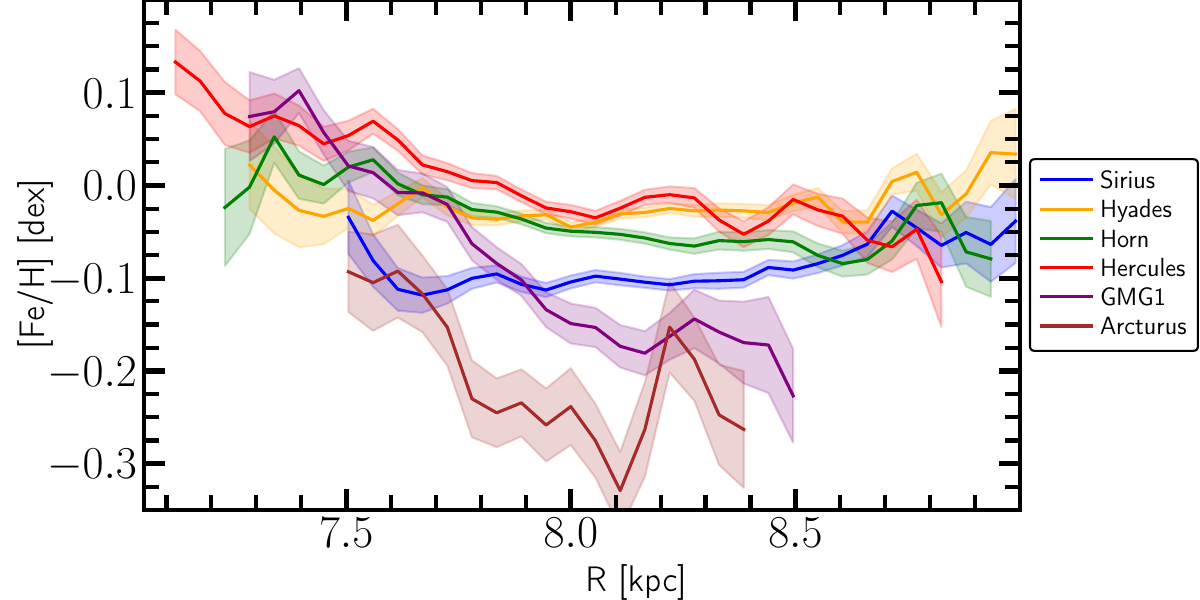}
	\caption[The median metallicity of stars within the selected moving groups with respect to the Galactocentric radius]{The median metallicity of stars within the selected moving groups with respect to the Galactocentric radius. The bin width in Galactocentric radius is $\SI{0.10}{kpc}$ and the bins are overlapping. The 1-$\sigma$ statistical uncertainty (shown in the shaded areas) on the mean trend lines is small, due to the large numbers of stars in each bin. We do not show bins containing fewer than 20 stars, which is why the mean relation does not always cover all radii. On average, stars scatter by $\SI{0.196}{dex}$ around their respective median lines.}
	\label{fig:stream_metal}
\end{figure*}

We show the average age and metallicity of each stream identified in \Sref{sec:identification_streams} as functions of Galactocentric radius in \Fref{fig:stream_age} and \Fref{fig:stream_metal}.
Sirius, Hyades, and Horn show roughly constant average ages and metallicities (with small trends for especially Horn) throughout the solar neighbourhood ($\lesssim\SI{1}{kpc}$ from the Sun), where the median age of Sirius, Hyades and Horn seems to be $\approx\SI{7.5}{Gyr}$ and metallicity between $-0.1 \textrm{ to }\SI{0.0}{dex}$. Our interpretation of this is that orbits of stars within these moving groups oscillate between the low and high $R$, and are at this point in time sufficiently mixed, which leads to an even distribution of ages and metallicities.

However, we observe a different behaviour for Hercules, GMG1, and Arcturus where ages and metallicities dependent more strongly on the Galactocentric radius. These moving groups were most likely created via the gravitational potential of the Galactic bar; either through a 4:1 OLR \citep[][]{Hunt2018} or stars captured in the Lagrange point $90^\circ$ off the major axis of the bar \citep{DOnghia2020}. Our own findings (\Fref{fig:stream_age} and \Fref{fig:stream_metal}) suggest that the stars in these moving groups did not have enough time to sufficiently mix, pointing towards a quickly evolving population that undergoes this migration process. This favours the Lagrange point hypothesis, as a Galactic bar that changes length on relatively short timescales would alter the position of the Lagrange point, sending distinct populations on their new orbits one after the other.

Another interesting aspect to note is that GMG1 \citep[first discovered and named in][]{Ramos2018} and Arcturus have lower metallicities than the other streams despite their angular momentum suggesting that stars in these moving groups would have originated from further inside the Galaxy (see \Fref{fig:stream_metal}). Their generally higher age around the solar neighbourhood could mean that it generally takes these stars a long time to reach their current orbit (with eccentricities high enough to reach $R=\SI{8}{kpc}$) on which we observe them, so these could be stars born relatively close to the Galactic Centre when the ISM in that region was not as chemically enriched as today. Hercules also displays this behaviour to a lesser degree with stars generally getting younger at $R<\SI{8}{kpc}$.
The weakness of a selection purely based on angular momentum becomes apparent here: as stars closer to the Galactic Centre would naturally have lower angular momenta and higher metallicities, it results in their identification as part of one of these moving groups with lower angular momentum. Naturally, we then see a high average metallicity for these moving groups in the medium age bin (as the number of medium aged stars in these moving groups is otherwise low, $n_\textrm{stars}\lesssim50$ in our sample which already counts these additional stars).


\subsubsection{Age--metallicity relation within moving groups}\label{sec:age_metal_relation}
\begin{figure*}
	\centering
	\includegraphics[width=0.85\textwidth]{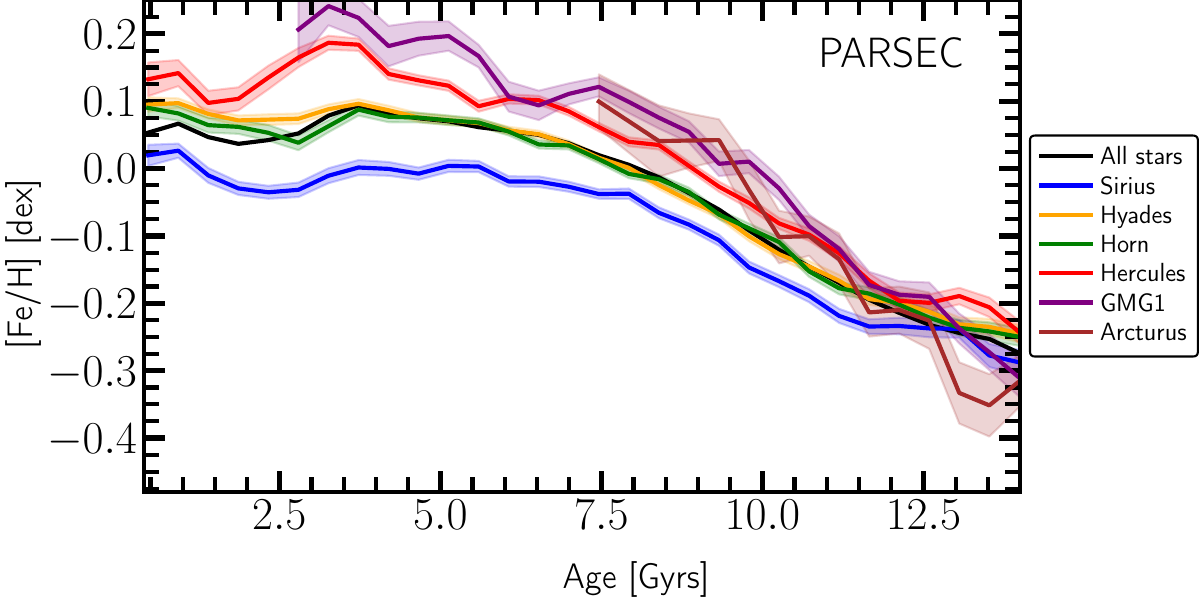}\\
    \vspace{0.3cm}
    \includegraphics[width=0.85\textwidth]{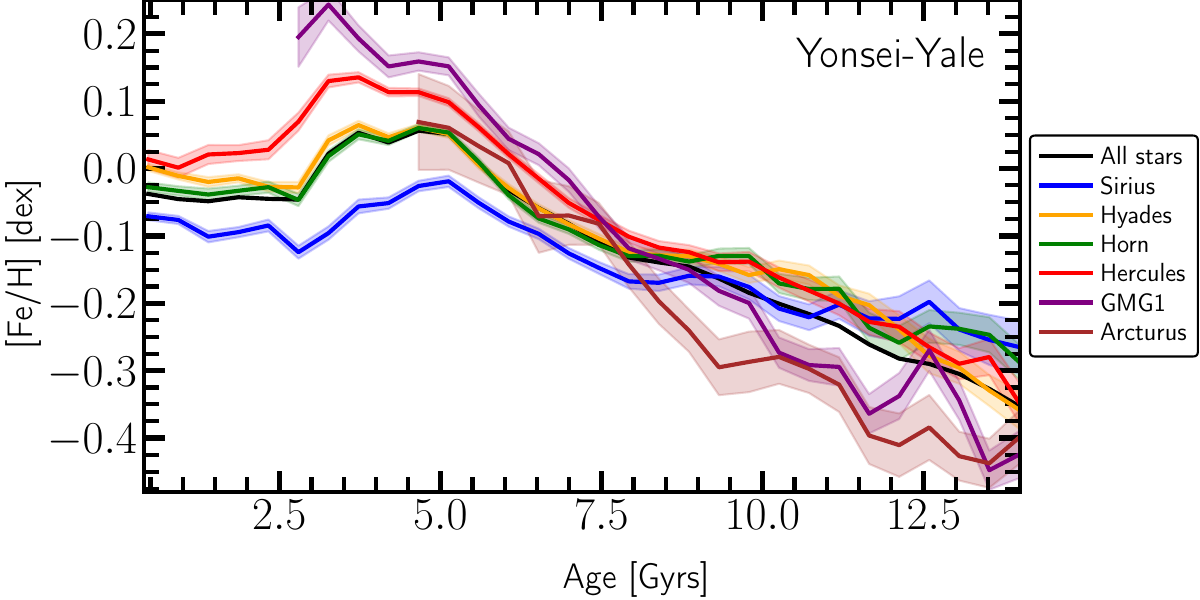}
	\caption[The median age--metallicity trends for stars within the selected moving groups]{The median age--metallicity trends for stars within the selected moving groups. The bin width in age is $\SI{0.8}{Gyr}$ and the bins are overlapping. The 1-$\sigma$ statistical uncertainty (shown in the shaded areas) on the mean trend lines is very small, due to the large numbers of stars in each bin. We do not show bins containing fewer than 20 stars, which is why the mean relation does not always cover across all ages. Top panel: ages from PARSEC-v2.0 isochrones. Bottom panel: ages from Yonsei-Yale isochrones \citep{Demarque2004}.
    }
	\label{fig:stream_age_metal}
\end{figure*}

In \Fref{fig:stream_age_metal}, we show the average metallicity of stream stars versus their age to see how it relates to e.g.\ the age--metallicity relation for our whole solar analogue sample (here shown as the black solid line). There are two panels with ages that were calculated with two sets of isochrones (PARSEC-v2.0 and Yonsei-Yale, \citealt{Demarque2004}), where we discuss the isochrone choice below (\Sref{sec:iso_choice}). The focus of this section is on the top diagram as we used PARSEC-v2.0 isochrones throughout this work but we show the Yonsei-Yale results to demonstrate that the main conclusions are independent of isochrone choice.

We can see that for older stars ($\textrm{age}\geq\SI{11}{Gyr}$) the age--metallicity relationship converges towards a shared trend for all moving groups. 
Our interpretation of this is that older stars have a sufficient amount of time to mix and interact with the Galactic environment so that, on average, all their initial structures (i.e.\ stellar clusters) have dissolved and the resulting distribution of old stars is the same throughout the Galaxy, as seen in simulations made by \citet{Loyola2015}.

For medium age stars ($\SI{6}{Gyr}\leq\textrm{age}<\SI{11}{Gyr}$) we can also observe that the metallicity increases with decreasing age in all moving groups. However, we observe that the metallicity increase is less steep for Sirius stars and steeper for Hercules, GMG1, and Arcturus stars. This is likely due to these moving groups containing mainly stars from further outside (in the case of Sirius) or inside (in the case of Hercules) the Galaxy, under the assumption that the Milky Way has a metallicity gradient with high metallicities in the centre and lower metallicities further outside \citep{Cheng2012, Jia2018}.
GMG1 and Arcturus show similar behaviour, with higher metallicities for younger stars, but the number of younger stars ($<\SI{7}{Gyr}$) in these moving groups is small ($<50$ for Arcturus). This suggests a certain limit on stellar migration, where time is required to move these young stars towards the solar neighbourhood. However, the young stars present seem to have high metallicities as expected for their low angular momentum as they were most likely born closer to the Galactic Centre.

Focusing on Sirius, Hyades, Horn and Hercules again, their sample of young stars (${\rm age}<\SI{6}{Gyr}$) seem to plateau in terms of metallicity (see \Fref{fig:stream_age_metal}). This is not surprising, as the increase of metallicity with decreasing age is not observed in the overall age--metallicity relationship (see \Fref{fig:age_metal_DR3}). Note that all of these moving groups seem to follow similar age--metallicity patterns but are offset from each other.
We also observe that the youngest stars have a lower average metallicity compared to stars aged $\approx4-\SI{5}{Gyr}$ within each of these moving groups. This drop in metallicity could suggest a strong in-fall of fresh gas \citep[see e.g.][]{Nissen2020} or simply that migration takes time to move metal-rich stars onto these orbits, as described above.
Our findings indicate that metal-rich young stars (${\rm [Fe/H]}>\SI{0.2}{dex}$ and ${\rm age}<\SI{3}{Gyr}$ in \Fref{fig:age_metal_DR3}) are missing rather than the entire population decreasing in metallicity, since we do observe significant numbers of young sub-solar metallicity ($\mathrm{[Fe/H]}<-0.1$) stars in our sample. 
Therefore, this favours the hypothesis of a lack of migration time over a major in-fall event (e.g.\ \citealt{Chen2019} suggested a migration timescale of $\SI{1}{kpc\,Gyr^{-1}}$). 

Furthermore, we can compare the age trends we see with the results of \citet{Sahlholdt2022}, where the population was split into different kinematic groups. GMG1 and Arcturus correspond to their population C while the other streams would probe the high (Sirius) and low (Hercules) $v_T$ ends of population B. We see a similar break in terms of metallicity between these two populations that were explored in their work, i.e. from Hercules to Sirius we have generally higher metallicities than for GMG1 and Arcturus. However, when we display these streams (which would be sub-populations of their population B and C) as we do in \Fref{fig:stream_age_metal}, we see that the transition from Sirius to Hercules and then to GMG1 and Arcturus is smoother than in \citet{Sahlholdt2022}. However, the overall age--metallicity trends seem to be similar.

\subsubsection{Isochrone choice}\label{sec:iso_choice}
Given the above results from the age-metallicity distribution, it is relevant to ensure that the choice of isochrones does not substantially impact our conclusions. Therefore, we can compare the two panels of \Fref{fig:stream_age_metal} where we can immediately see that the ages from PARSEC-v2.0 with \SAMD\ allow over a larger range (up to $\SI{20}{Gyr}$) than Yonsei-Yale with q$^2$, which leads to the age axis for the latter appearing more contracted. However, the main features that were discussed in \Sref{sec:age_metal_relation} are still present, e.g.\ the old population having the same median age independent of moving group and the metallicity peaking at a medium age and falling towards the youngest stars. Overall, while we favour the more up-to-date PARSEC isochrones to deliver reliable ages for our sample, it is reassuring to find similar results independent of the isochrones used.

\section{Conclusion}\label{sec:conclusion_ch4}
We analysed GALAH DR3 spectra using $\EPIC$ \citep{CLehmann2022} and \SAMD\ \citep{Sahlholdt2020} with the goal to test the origins of moving groups using solar analogues as probes. The Hercules stream in particular is well understood to originate closer to the Galactic Centre \citep{Hunt2018} with stars within the stream then migrating towards the solar neighbourhood.
We identified $72{,}288$ solar analogue targets in GALAH DR3 and derived their ages using isochrone fitting (\SAMD).
We compared the age uncertainties of this approach to those of the GALAH BSTEP method and showed that our $\EPIC$/\SAMD\ approach has lower uncertainties by a factor of $\approx1.5$. 
Using a spectrograph that provides a larger spectral range, and therefore more possible absorption features, could lead to even lower age uncertainties.

Using a large number of solar analogue stars with precise stellar parameter measurements (see \Fref{fig:uncertainties_DR3}), we investigated potential trends in moving groups of the solar neighbourhood. We identified stars as part of moving groups via their angular momentum with the moving groups in this work being Sirius, Hyades, Horn, Hercules, GMG1 and Arcturus. There are several notable a few differences within the moving groups: 
(i) GMG1 and Arcturus are older and more metal-poor than the other moving groups. This could mean that they consist of a mainly older population that had time to migrate outwards via a blurring process while younger stars maybe did not have the necessary time.
(ii) Hercules, GMG1, and Arcturus stand out as moving groups with stars that vary in both age and metallicity throughout their radial extent. They have stars with increasing metallicity and decreasing age closer to the Galactic Centre, while each of the other moving groups having relatively constant distributions of age and metallicity throughout their radial extent (with the exception of Horn, which shows a smaller scale of variation). This might be because these moving groups are not well mixed, which means new stars are being fed into them closer to the Galactic centre, possibly due to resonance effects of the Galactic bar.
(iii) Within each individual stream, the age--metallicity behaviour of stars is similar, with a rise in metallicity from old to young stars and a plateau metallicity that each stream reaches for stars $3.5-\SI{5}{Gyr}$ and younger (excluding GMG1 and Arcturus for which not enough young stars were identified in our sample). This constant metallicity level is different for each stream, with Sirius around $\SI{-0.02}{dex}$, Hyades and Horn around $\SI{0.05}{dex}$ and Hercules at $\SI{0.15}{dex}$. 
The overall age-metallicity distribution of the sample (black line in \Fref{fig:stream_age_metal}) and individual moving groups also show a slight drop in metallicity for stars with ages $\sim\SI{3.5}{Gyr}$ and younger, which suggests that the metal-rich young stars did not have enough time to migrate outwards and we are left with stars that were predominantly locally born and therefore more metal-poor. The over-density of medium-aged metal-rich stars in \Fref{fig:age_metal_DR3} and the `missing' young metal-rich stars are consistent with a slow migration process \citep[e.g.][]{Chen2019} of order $\SI{1}{kpc\,Gyr^{-1}}$, insufficient to populate the young and metal-rich stars in the available time.

\section*{Acknowledgements}
We would like both Michael Hayden and Ivan Minchev for sharing their respective knowledge on stellar migration, which helped to get this project going. We would also like to thank Diane Feuillet for giving critique at the later stages of this project and Heitor Ernandes for sharing his expertise with the \SAMD\ code.

CL, MTM and FL acknowledge the support of the Australian Research Council through \textsl{Future Fellowship} grant FT180100194.

This work made use of the Third Data Release of the GALAH Survey \citep{Buder2021}. The GALAH Survey is based on data acquired through the Australian Astronomical Observatory, under programs: A/2013B/13 (The GALAH pilot survey); A/2014A/25, A/2015A/19, A2017A/18 (The GALAH survey phase 1); A2018A/18 (Open clusters with HERMES); A2019A/1 (Hierarchical star formation in Ori OB1); A2019A/15 (The GALAH survey phase 2); A/2015B/19, A/2016A/22, A/2016B/10, A/2017B/16, A/2018B/15 (The HERMES-TESS program); and A/2015A/3, A/2015B/1, A/2015B/19, A/2016A/22, A/2016B/12, A/2017A/14 (The HERMES K2-follow-up program). We acknowledge the traditional owners of the land on which the AAT stands, the Gamilaraay people, and pay our respects to elders past and present. This paper includes data that has been provided by AAO Data Central (datacentral.org.au).

\section*{Data Availability}
The value added catalogue for the stars used in this paper can be found in the supplementary material and is available at CDS via anonymous ftp to cdsarc.u-strasbg.fr (130.79.128.5) or via \url{http://vizier.cds.unistra.fr/viz-bin/VizieR?-source=J/MNRAS/536/498}.



\bibliographystyle{mnras}
\bibliography{Bibliography} 

\begin{thebibliography}{}
\makeatletter
\relax
\def\mn@urlcharsother{\let\do\@makeother \do\$\do\&\do\#\do\^\do\_\do\%\do\~}
\def\mn@doi{\begingroup\mn@urlcharsother \@ifnextchar [ {\mn@doi@}
  {\mn@doi@[]}}
\def\mn@doi@[#1]#2{\def\@tempa{#1}\ifx\@tempa\@empty \href
  {http://dx.doi.org/#2} {doi:#2}\else \href {http://dx.doi.org/#2} {#1}\fi
  \endgroup}
\def\mn@eprint#1#2{\mn@eprint@#1:#2::\@nil}
\def\mn@eprint@arXiv#1{\href {http://arxiv.org/abs/#1} {{\tt arXiv:#1}}}
\def\mn@eprint@dblp#1{\href {http://dblp.uni-trier.de/rec/bibtex/#1.xml}
  {dblp:#1}}
\def\mn@eprint@#1:#2:#3:#4\@nil{\def\@tempa {#1}\def\@tempb {#2}\def\@tempc
  {#3}\ifx \@tempc \@empty \let \@tempc \@tempb \let \@tempb \@tempa \fi \ifx
  \@tempb \@empty \def\@tempb {arXiv}\fi \@ifundefined
  {mn@eprint@\@tempb}{\@tempb:\@tempc}{\expandafter \expandafter \csname
  mn@eprint@\@tempb\endcsname \expandafter{\@tempc}}}

\bibitem[\protect\citeauthoryear{Antoja, Figueras, Torra, Valenzuela  \&
  Pichardo}{Antoja et~al.}{2010}]{Antoja2010}
Antoja T.,  Figueras F.,  Torra J.,  Valenzuela O.,   Pichardo B.,  2010, in ,
  Vol.~4, Lecture Notes and Essays in Astrophysics.
Spanish Royal Physical Society (RSEF), pp 13--31, \url
  {https://ui.adsabs.harvard.edu/abs/2010LNEA....4...13B}

\bibitem[\protect\citeauthoryear{Bailer-Jones, Rybizki, Fouesneau, Mantelet  \&
  Andrae}{Bailer-Jones et~al.}{2018}]{BailerJones2018}
Bailer-Jones C. A.~L.,  Rybizki J.,  Fouesneau M.,  Mantelet G.,   Andrae R.,
  2018, \mn@doi [ApJ] {10.3847/1538-3881/aacb21}, 156, 58

\bibitem[\protect\citeauthoryear{Bailer-Jones, Rybizki, Fouesneau, Demleitner
  \& Andrae}{Bailer-Jones et~al.}{2021}]{BailerJones2021}
Bailer-Jones C. A.~L.,  Rybizki J.,  Fouesneau M.,  Demleitner M.,   Andrae R.,
   2021, \mn@doi [ApJ] {10.3847/1538-3881/abd806}, 161, 147

\bibitem[\protect\citeauthoryear{Basu \& Kinnane}{Basu \&
  Kinnane}{2018}]{Basu2018}
Basu S.,  Kinnane A.,  2018, \mn@doi [ApJ] {10.3847/1538-4357/aae922}, 869, 8

\bibitem[\protect\citeauthoryear{Bensby, Feltzing  \& Oey}{Bensby
  et~al.}{2014}]{Bensby2014}
Bensby T.,  Feltzing S.,   Oey M.~S.,  2014, \mn@doi [A{\&}A]
  {10.1051/0004-6361/201322631}, 562, A71

\bibitem[\protect\citeauthoryear{Bonanno, Schlattl  \& Patern{\`{o}}}{Bonanno
  et~al.}{2002}]{Bonanno2002}
Bonanno A.,  Schlattl H.,   Patern{\`{o}} L.,  2002, \mn@doi [A{\&}A]
  {10.1051/0004-6361:20020749}, 390, 1115

\bibitem[\protect\citeauthoryear{Bovy}{Bovy}{2015}]{galpy15}
Bovy J.,  2015, \mn@doi [ApJS] {10.1088/0067-0049/216/2/29}, 216, 29

\bibitem[\protect\citeauthoryear{Buder et~al.,}{Buder et~al.}{2021}]{Buder2021}
Buder S.,  et~al., 2021, \mn@doi [MNRAS] {10.1093/mnras/stab1242}, 506, 150

\bibitem[\protect\citeauthoryear{Casali et~al.,}{Casali
  et~al.}{2020}]{Casali2020}
Casali G.,  et~al., 2020, \mn@doi [A\&A] {10.1051/0004-6361/202038055}, 639,
  A127

\bibitem[\protect\citeauthoryear{Chen, Bressan, Girardi, Marigo, Kong  \&
  Lanza}{Chen et~al.}{2015}]{Chen2015}
Chen Y.,  Bressan A.,  Girardi L.,  Marigo P.,  Kong X.,   Lanza A.,  2015,
  \mn@doi [MNRAS] {10.1093/mnras/stv1281}, 452, 1068

\bibitem[\protect\citeauthoryear{Chen, Zhao, Zhao, Liang, Wu, Jia, Tian  \&
  Liu}{Chen et~al.}{2019}]{Chen2019}
Chen Y.~Q.,  Zhao G.,  Zhao J.~K.,  Liang X.~L.,  Wu Y.~Q.,  Jia Y.~P.,  Tian
  H.,   Liu J.~M.,  2019, \mn@doi [ApJ] {10.3847/1538-3881/ab5283}, 158, 249

\bibitem[\protect\citeauthoryear{Cheng et~al.,}{Cheng et~al.}{2012}]{Cheng2012}
Cheng J.~Y.,  et~al., 2012, \mn@doi [ApJ] {10.1088/0004-637x/746/2/149}, 746,
  149

\bibitem[\protect\citeauthoryear{Connelly, Bizzarro, Krot, Nordlund, Wielandt
  \& Ivanova}{Connelly et~al.}{2012}]{Connelly2012}
Connelly J.~N.,  Bizzarro M.,  Krot A.~N.,  Nordlund {\AA}.,  Wielandt D.,
  Ivanova M.~A.,  2012, \mn@doi [Science] {10.1126/science.1226919}, 338, 651

\bibitem[\protect\citeauthoryear{Cowan, Pfeiffer, Kratz, Thielemann, Sneden,
  Burles, Tytler  \& Beers}{Cowan et~al.}{1999}]{Cowan1999}
Cowan J.~J.,  Pfeiffer B.,  Kratz K.-L.,  Thielemann F.-K.,  Sneden C.,  Burles
  S.,  Tytler D.,   Beers T.~C.,  1999, \mn@doi [ApJ] {10.1086/307512}, 521,
  194

\bibitem[\protect\citeauthoryear{D'Onghia \& Aguerri}{D'Onghia \&
  Aguerri}{2020}]{DOnghia2020}
D'Onghia E.,  Aguerri J. A.~L.,  2020, \mn@doi [ApJ]
  {10.3847/1538-4357/ab6bd6}, 890, 117

\bibitem[\protect\citeauthoryear{Dehnen}{Dehnen}{2000}]{Dehnen2000}
Dehnen W.,  2000, \mn@doi [ApJ] {10.1086/301226}, 119, 800

\bibitem[\protect\citeauthoryear{Demarque \& Larson}{Demarque \&
  Larson}{1964}]{Demarque1964}
Demarque P.~R.,  Larson R.~B.,  1964, \mn@doi [\apj] {10.1086/147948}, 140, 544

\bibitem[\protect\citeauthoryear{Demarque, Woo, Kim  \& Yi}{Demarque
  et~al.}{2004}]{Demarque2004}
Demarque P.,  Woo J.-H.,  Kim Y.-C.,   Yi S.~K.,  2004, \mn@doi [ApJS]
  {10.1086/424966}, 155, 667

\bibitem[\protect\citeauthoryear{Eggen}{Eggen}{1995}]{Eggen1995}
Eggen O.~J.,  1995, \mn@doi [ApJ] {10.1086/117567}, 110, 823

\bibitem[\protect\citeauthoryear{Eggen}{Eggen}{1996}]{Eggen1996}
Eggen O.~J.,  1996, \mn@doi [AJ] {10.1086/117901}, 111, 1615

\bibitem[\protect\citeauthoryear{Famaey, Jorissen, Luri, Mayor, Udry, Dejonghe
  \& Turon}{Famaey et~al.}{2005}]{Famaey2005}
Famaey B.,  Jorissen A.,  Luri X.,  Mayor M.,  Udry S.,  Dejonghe H.,   Turon
  C.,  2005, \mn@doi [A\&A] {10.1051/0004-6361:20041272}, 430, 165

\bibitem[\protect\citeauthoryear{Famaey, Pont, Luri, Udry, Mayor  \&
  Jorissen}{Famaey et~al.}{2007}]{Famaey2007}
Famaey B.,  Pont F.,  Luri X.,  Udry S.,  Mayor M.,   Jorissen A.,  2007,
  \mn@doi [A\&A] {10.1051/0004-6361:20065706}, 461, 957

\bibitem[\protect\citeauthoryear{Fragkoudi et~al.,}{Fragkoudi
  et~al.}{2019}]{Fragkoudi2019}
Fragkoudi F.,  et~al., 2019, \mn@doi [MNRAS] {10.1093/mnras/stz1875}

\bibitem[\protect\citeauthoryear{Frebel \& Kratz}{Frebel \&
  Kratz}{2008}]{Frebel2008}
Frebel A.,  Kratz K.-L.,  2008, \mn@doi [Proceedings of the International
  Astronomical Union] {10.1017/s1743921309032104}, 4, 449

\bibitem[\protect\citeauthoryear{Freeman \& Bland-Hawthorn}{Freeman \&
  Bland-Hawthorn}{2002}]{Freeman2002}
Freeman K.,  Bland-Hawthorn J.,  2002, \mn@doi [ARA\&A]
  {10.1146/annurev.astro.40.060401.093840}, 40, 487

\bibitem[\protect\citeauthoryear{{GRAVITY Collaboration} et~al.,}{{GRAVITY
  Collaboration} et~al.}{2019}]{Abuter2019}
{GRAVITY Collaboration} et~al., 2019, \mn@doi [A{\&}A]
  {10.1051/0004-6361/201935656}, 625, L10

\bibitem[\protect\citeauthoryear{{Gaia Collaboration} et~al.,}{{Gaia
  Collaboration} et~al.}{2016}]{Collaboration2016}
{Gaia Collaboration} et~al., 2016, \mn@doi [A\&A]
  {10.1051/0004-6361/201629272}, 595, A1

\bibitem[\protect\citeauthoryear{{Gaia Collaboration} et~al.,}{{Gaia
  Collaboration} et~al.}{2018}]{Collaboration2018}
{Gaia Collaboration} et~al., 2018, \mn@doi [A{\&}A]
  {10.1051/0004-6361/201833051}, 616, A1

\bibitem[\protect\citeauthoryear{{Gaia Collaboration} et~al.,}{{Gaia
  Collaboration} et~al.}{2021}]{Collaboration2021}
{Gaia Collaboration} et~al., 2021, \mn@doi [A{\&}A]
  {10.1051/0004-6361/202039657}, 649, A1

\bibitem[\protect\citeauthoryear{{Gaia Collaboration} et~al.,}{{Gaia
  Collaboration} et~al.}{2023a}]{Collaboration2023}
{Gaia Collaboration} et~al., 2023a, \mn@doi [A\&A]
  {10.1051/0004-6361/202243940}, 674, A1

\bibitem[\protect\citeauthoryear{{Gaia Collaboration} et~al.,}{{Gaia
  Collaboration} et~al.}{2023b}]{GaiaCollaboration2022}
{Gaia Collaboration} et~al., 2023b, \mn@doi [A\&A]
  {10.1051/0004-6361/202243800}, 674, A39

\bibitem[\protect\citeauthoryear{Gardner \& Flynn}{Gardner \&
  Flynn}{2010a}]{Gardner2010}
Gardner E.,  Flynn C.,  2010a, \mn@doi [MNRAS]
  {10.1111/j.1365-2966.2010.16470.x}, 405, 545

\bibitem[\protect\citeauthoryear{Gardner \& Flynn}{Gardner \&
  Flynn}{2010b}]{Gardner2010a}
Gardner E.,  Flynn C.,  2010b, \mn@doi [MNRAS]
  {10.1111/j.1365-2966.2010.16913.x}, 406, 701

\bibitem[\protect\citeauthoryear{Hayden et~al.,}{Hayden
  et~al.}{2022}]{Hayden2022}
Hayden M.~R.,  et~al., 2022, \mn@doi [MNRAS] {10.1093/mnras/stac2787}, 517,
  5325

\bibitem[\protect\citeauthoryear{Hunt \& Bovy}{Hunt \& Bovy}{2018}]{Hunt2018}
Hunt J. A.~S.,  Bovy J.,  2018, \mn@doi [MNRAS] {10.1093/mnras/sty921}, 477,
  3945

\bibitem[\protect\citeauthoryear{Jia, Chen, Zhao, Xue, Zhao, Yang  \& Li}{Jia
  et~al.}{2018}]{Jia2018}
Jia Y.,  Chen Y.,  Zhao G.,  Xue X.,  Zhao J.,  Yang C.,   Li C.,  2018,
  \mn@doi [ApJ] {10.3847/1538-4357/aad3bb}, 863, 93

\bibitem[\protect\citeauthoryear{J{\o}rgensen \& Lindegren}{J{\o}rgensen \&
  Lindegren}{2005}]{Joergensen2005}
J{\o}rgensen B.~R.,  Lindegren L.,  2005, \mn@doi [A{\&}A]
  {10.1051/0004-6361:20042185}, 436, 127

\bibitem[\protect\citeauthoryear{Khanna et~al.,}{Khanna
  et~al.}{2019}]{Khanna2019}
Khanna S.,  et~al., 2019, \mn@doi [MNRAS] {10.1093/mnras/stz2462}, 489, 4962

\bibitem[\protect\citeauthoryear{Lehmann, Murphy, Liu, Flynn  \& Berke}{Lehmann
  et~al.}{2022}]{CLehmann2022}
Lehmann C.,  Murphy M.~T.,  Liu F.,  Flynn C.,   Berke D.~A.,  2022, \mn@doi
  [MNRAS] {10.1093/mnras/stac421}, 512, 11

\bibitem[\protect\citeauthoryear{Lehmann, Murphy, Liu, Flynn, Smith  \&
  Berke}{Lehmann et~al.}{2023}]{Lehmann2023}
Lehmann C.,  Murphy M.~T.,  Liu F.,  Flynn C.,  Smith D.,   Berke D.~A.,  2023,
  \mn@doi [MNRAS] {10.1093/mnras/stad381}, 521, 148

\bibitem[\protect\citeauthoryear{Loebman, Ro{\v{s}}kar, Debattista,
  Ivezi{\'{c}}, Quinn  \& Wadsley}{Loebman et~al.}{2011}]{Loebman2011}
Loebman S.~R.,  Ro{\v{s}}kar R.,  Debattista V.~P.,  Ivezi{\'{c}} {\v{Z}}.,
  Quinn T.~R.,   Wadsley J.,  2011, \mn@doi [ApJ] {10.1088/0004-637x/737/1/8},
  737, 8

\bibitem[\protect\citeauthoryear{Loyola, Flynn, Hurley  \& Gibson}{Loyola
  et~al.}{2015}]{Loyola2015}
Loyola G. R. I.~M.,  Flynn C.,  Hurley J.~R.,   Gibson B.~K.,  2015, \mn@doi
  [MNRAS] {10.1093/mnras/stv550}, 449, 4443

\bibitem[\protect\citeauthoryear{Lucchini, Pellett, D'Onghia  \&
  Aguerri}{Lucchini et~al.}{2022}]{Lucchini2022}
Lucchini S.,  Pellett E.,  D'Onghia E.,   Aguerri J. A.~L.,  2022, \mn@doi
  [MNRAS] {10.1093/mnras/stac3519}, 519, 432

\bibitem[\protect\citeauthoryear{Minchev, Chiappini  \& Martig}{Minchev
  et~al.}{2013}]{Minchev2013}
Minchev I.,  Chiappini C.,   Martig M.,  2013, \mn@doi [A\&A]
  {10.1051/0004-6361/201220189}, 558, A9

\bibitem[\protect\citeauthoryear{Minchev, Martig, Streich, Scannapieco, de Jong
   \& Steinmetz}{Minchev et~al.}{2015}]{Minchev2015}
Minchev I.,  Martig M.,  Streich D.,  Scannapieco C.,  de Jong R.~S.,
  Steinmetz M.,  2015, \mn@doi [ApJ] {10.1088/2041-8205/804/1/l9}, 804, L9

\bibitem[\protect\citeauthoryear{Minchev et~al.,}{Minchev
  et~al.}{2018}]{Minchev2018}
Minchev I.,  et~al., 2018, \mn@doi [MNRAS] {10.1093/mnras/sty2033}, 481, 1645

\bibitem[\protect\citeauthoryear{Mints \& Hekker}{Mints \&
  Hekker}{2017}]{Mints2017}
Mints A.,  Hekker S.,  2017, \mn@doi [A\&A] {10.1051/0004-6361/201630090}, 604,
  A108

\bibitem[\protect\citeauthoryear{Monari, Famaey, Siebert, Duchateau,
  Lorscheider  \& Bienaym{\'{e}}}{Monari et~al.}{2016}]{Monari2016}
Monari G.,  Famaey B.,  Siebert A.,  Duchateau A.,  Lorscheider T.,
  Bienaym{\'{e}} O.,  2016, \mn@doi [MNRAS] {10.1093/mnras/stw2807}, 465, 1443

\bibitem[\protect\citeauthoryear{Moya, Sarro, Delgado-Mena, Chaplin, Adibekyan
  \& Blanco-Cuaresma}{Moya et~al.}{2022}]{Moya2022}
Moya A.,  Sarro L.~M.,  Delgado-Mena E.,  Chaplin W.~J.,  Adibekyan V.,
  Blanco-Cuaresma S.,  2022, \mn@doi [A{\&}A] {10.1051/0004-6361/202141125},
  660, A15

\bibitem[\protect\citeauthoryear{Nguyen et~al.,}{Nguyen
  et~al.}{2022}]{Nguyen2022}
Nguyen C.~T.,  et~al., 2022, \mn@doi [A\&A] {10.1051/0004-6361/202244166}, 665,
  A126

\bibitem[\protect\citeauthoryear{Nissen}{Nissen}{2016}]{Nissen2016}
Nissen P.~E.,  2016, \mn@doi [A\&A] {10.1051/0004-6361/201628888}, 593, A65

\bibitem[\protect\citeauthoryear{Nissen, Christensen-Dalsgaard, Mosumgaard,
  Aguirre, Spitoni  \& Verma}{Nissen et~al.}{2020}]{Nissen2020}
Nissen P.~E.,  Christensen-Dalsgaard J.,  Mosumgaard J.~R.,  Aguirre V.~S.,
  Spitoni E.,   Verma K.,  2020, \mn@doi [A\&A] {10.1051/0004-6361/202038300},
  640, A81

\bibitem[\protect\citeauthoryear{P{\'e}rez-Villegas, Portail, Wegg  \&
  Gerhard}{P{\'e}rez-Villegas et~al.}{2017}]{PerezVillegas2017}
P{\'e}rez-Villegas A.,  Portail M.,  Wegg C.,   Gerhard O.,  2017, \mn@doi
  [ApJL] {10.3847/2041-8213/aa6c26}, 840, L2

\bibitem[\protect\citeauthoryear{Piskunov \& Valenti}{Piskunov \&
  Valenti}{2017}]{Piskunov2017}
Piskunov N.,  Valenti J.~A.,  2017, \mn@doi [A{\&}A]
  {10.1051/0004-6361/201629124}, 597, A16

\bibitem[\protect\citeauthoryear{Proctor}{Proctor}{1869}]{Proctor1869}
Proctor R.~A.,  1869, \mn@doi [MNRAS] {10.1093/mnras/30.2.50}, 30, 50

\bibitem[\protect\citeauthoryear{Pr{\v{s}}a et~al.,}{Pr{\v{s}}a
  et~al.}{2016}]{Prsa2016}
Pr{\v{s}}a A.,  et~al., 2016, \mn@doi [ApJ] {10.3847/0004-6256/152/2/41}, 152,
  41

\bibitem[\protect\citeauthoryear{Ram{\'{\i}}rez et~al.,}{Ram{\'{\i}}rez
  et~al.}{2014}]{Ramirez2014}
Ram{\'{\i}}rez I.,  et~al., 2014, \mn@doi [A{\&}A]
  {10.1051/0004-6361/201424244}, 572, A48

\bibitem[\protect\citeauthoryear{Ramos, Antoja  \& Figueras}{Ramos
  et~al.}{2018}]{Ramos2018}
Ramos P.,  Antoja T.,   Figueras F.,  2018, \mn@doi [A{\&}A]
  {10.1051/0004-6361/201833494}, 619, A72

\bibitem[\protect\citeauthoryear{Rocha, Almeida, Damineli, Navarete,
  Abdul-Masih  \& Mace}{Rocha et~al.}{2022}]{Rocha2022}
Rocha D.~F.,  Almeida L.~A.,  Damineli A.,  Navarete F.,  Abdul-Masih M.,
  Mace G.~N.,  2022, \mn@doi [MNRAS] {10.1093/mnras/stac2927}, 517, 3749

\bibitem[\protect\citeauthoryear{Rottensteiner \& Meingast}{Rottensteiner \&
  Meingast}{2024}]{Rottensteiner2024}
Rottensteiner A.,  Meingast S.,  2024, \mn@doi [A\&A]
  {10.1051/0004-6361/202347701}, 690, A16

\bibitem[\protect\citeauthoryear{Sahlholdt}{Sahlholdt}{2020}]{Sahlholdt2020}
Sahlholdt C.,  2020, csahlholdt/SAMD: First SAMD release, GitHub,
  \mn@doi{10.5281/zenodo.3941733}, \url
  {https://ui.adsabs.harvard.edu/abs/2020zndo...3941733S}

\bibitem[\protect\citeauthoryear{Sahlholdt \& Lindegren}{Sahlholdt \&
  Lindegren}{2021}]{Sahlholdt2021}
Sahlholdt C.~L.,  Lindegren L.,  2021, \mn@doi [MNRAS] {10.1093/mnras/stab034},
  502, 845

\bibitem[\protect\citeauthoryear{Sahlholdt, Feltzing  \& Feuillet}{Sahlholdt
  et~al.}{2022}]{Sahlholdt2022}
Sahlholdt C.~L.,  Feltzing S.,   Feuillet D.~K.,  2022, \mn@doi [MNRAS]
  {10.1093/mnras/stab3681}, 510, 4669

\bibitem[\protect\citeauthoryear{Sanders \& Das}{Sanders \&
  Das}{2018}]{Sanders2018}
Sanders J.~L.,  Das P.,  2018, \mn@doi [MNRAS] {10.1093/mnras/sty2490}, 481,
  4093

\bibitem[\protect\citeauthoryear{Schlegel, Finkbeiner  \& Davis}{Schlegel
  et~al.}{1998}]{Schlegel1998}
Schlegel D.~J.,  Finkbeiner D.~P.,   Davis M.,  1998, \mn@doi [ApJ]
  {10.1086/305772}, 500, 525

\bibitem[\protect\citeauthoryear{Sch{\"o}nrich \& Binney}{Sch{\"o}nrich \&
  Binney}{2009}]{Schoenrich2009}
Sch{\"o}nrich R.,  Binney J.,  2009, \mn@doi [MNRAS]
  {10.1111/j.1365-2966.2009.15365.x}, 399, 1145

\bibitem[\protect\citeauthoryear{Sellwood}{Sellwood}{2014}]{Sellwood2014}
Sellwood J.~A.,  2014, \mn@doi [RMP] {10.1103/revmodphys.86.1}, 86, 1

\bibitem[\protect\citeauthoryear{Shah et~al.,}{Shah et~al.}{2023}]{Shah2023}
Shah S.~P.,  et~al., 2023, \mn@doi [ApJ] {10.3847/1538-4357/acb8af}, 948, 122

\bibitem[\protect\citeauthoryear{Sharma et~al.,}{Sharma
  et~al.}{2017}]{Sharma2017}
Sharma S.,  et~al., 2017, \mn@doi [MNRAS] {10.1093/mnras/stx2582}, 473, 2004

\bibitem[\protect\citeauthoryear{{Silva Aguirre} et~al.,}{{Silva Aguirre}
  et~al.}{2018}]{Aguirre2018}
{Silva Aguirre} V.,  et~al., 2018, \mn@doi [MNRAS] {10.1093/mnras/sty150}

\bibitem[\protect\citeauthoryear{Skuljan, Hearnshaw  \& Cottrell}{Skuljan
  et~al.}{1999}]{Skuljan1999}
Skuljan J.,  Hearnshaw J.~B.,   Cottrell P.~L.,  1999, \mn@doi [MNRAS]
  {10.1046/j.1365-8711.1999.02736.x}, 308, 731

\bibitem[\protect\citeauthoryear{Valentini et~al.,}{Valentini
  et~al.}{2019}]{Valentini2019}
Valentini M.,  et~al., 2019, \mn@doi [A{\&}A] {10.1051/0004-6361/201834081},
  627, A173

\bibitem[\protect\citeauthoryear{{Viscasillas V{\'{a}}zquez}
  et~al.,}{{Viscasillas V{\'{a}}zquez} et~al.}{2022}]{Vazquez2022}
{Viscasillas V{\'{a}}zquez} C.,  et~al., 2022, \mn@doi [A{\&}A]
  {10.1051/0004-6361/202142937}, 660, A135

\bibitem[\protect\citeauthoryear{Willett et~al.,}{Willett
  et~al.}{2023}]{Willett2023}
Willett E.,  et~al., 2023, \mn@doi [MNRAS] {10.1093/mnras/stad2374}, 526, 2141

\bibitem[\protect\citeauthoryear{Zwitter et~al.,}{Zwitter
  et~al.}{2018}]{Zwitter2018}
Zwitter T.,  et~al., 2018, \mn@doi [MNRAS] {10.1093/mnras/sty2293}, 481, 645

\makeatother
\end{thebibliography}




\appendix


\bsp	
\label{lastpage}
\end{CJK*}
\end{document}